\documentclass[english, 9pt,  a4paper, onecolumn, technote]{IEEEconf}

\usepackage[T1]{fontenc}
\usepackage{times}

\usepackage{prettyref}
\usepackage{wrapfig}
\usepackage{amsmath}
\usepackage{graphicx}

\usepackage{url}
\usepackage{babel}

\begin{document}

\title{A Characterization of the SPARC T3-4 System}

\author{Michiel W. van Tol\\
\begin{affiliation}
Computer Systems Architecture group, Institute for Informatics\\
Faculty of Science, University of Amsterdam\\
Amsterdam, The Netherlands
\end{affiliation} \\
\email{mwvantol@uva.nl}
}

\maketitle

\begin{abstract}
This technical report covers a set of experiments on the 64-core SPARC T3-4 system, comparing it to two similar AMD and Intel systems. Key characteristics as maximum integer and floating point arithmetic throughput are measured as well as memory throughput, showing the scalability of the SPARC T3-4 system. The performance of POSIX threads primitives is characterized and compared in detail, such as thread creation and mutex synchronization. Scalability tests with a fine grained multithreaded runtime are performed, showing problems with atomic CAS operations on such physically highly parallel systems.
\end{abstract}

\pagebreak{}
\tableofcontents{}
\pagebreak{}

\section{Introduction}

This document is a technical report discussing tests and results on a Sun/Oracle SPARC T3-4 system that the Computer Systems Architecture (CSA) group at the University of Amsterdam had access to through the CMT beta program.  The experiments described in this document were solely conducted by the author, and do not provide a full account of all tests done on the T3-4 by the CSA group within the test period. The opinions expressed in this document are solely those of the author, and are not official opinions of neither the CSA group nor the University of Amsterdam.

The T3-4 system is a 4 way multicore SPARC~T3~\cite{T3core2010} system with the CPUs running at 1.67~GHz. Each CPU contains 16~cores, and each core contains 8 hardware thread contexts, for an aggregated total of 512~hardware threads in the whole system. For a more detailed description of the architecture and organization of the system, we refer to the available literature~\cite{T3core2010, T3whitepaper2010, SPARC-T3, T3-4System}.

The rest of this document is structured as follows; first, we will
discuss the background of the research the author is involved in within
the CSA group, to show from which standpoint these experiments were
selected. Then we will thoroughly discuss each individual experiment;
how it was defined, analyze its results, and when applicable, compare
the test results to two reference systems. After all experiments have
been discussed, we draw our overall conclusion about the T3-4 system.
The two reference systems that we used consist of a Sun/Oracle X4470
server~\cite{X4470} fitted with four 6-core Intel Xeon E7530~\cite{Xeon7500}
processors running at 1.87~GHz, and a Dell PowerEdge R815~\cite{DellR815}
AMD \emph{Magny Cours} system which has four 12-core AMD Opteron
6172~\cite{Opteron6000} processors running at 2.1 GHz. Both systems
are running CentOS 5.5 64-bit Linux, kernel version 2.6.18. As the
Xeon E7530 processor has HyperThreading, giving us two threads per
core, both reference systems present themselves through the OS as
48-cpu machines. Throughout this document we will refer to these machines
as \emph{X4470} and \emph{Magny Cours}. The T3-4 system is running
Oracle Solaris~10~9/10.

\section{Background}\label{sec:Background}

As the development of performance using single cores has stagnated
due to the power wall, a switch has been made in the last few years
towards multi- and many-core architectures. In order to exploit the
potential of architectures with hundreds, thousands or more cores,
concurrency should be the norm and starting many parallel tasks should
be cheap. This is reflected in one of the areas the CSA group is doing
its research in, where we are doing work on the design of a conceptual many-core
architecture; the Microgrid \cite{Microgrid09}. This CMP architecture
employs fine grained data driven scheduling between many small threads
of execution, a core can accommodate up to hundreds of threads and
switch between them on a cycle-to-cycle basis, allowing for fine grained
latency hiding. These threads, so called Microthreads \cite{Microthread06},
can be created cheaply by the core's logic in a few cycles, and can
consist of anything ranging from a few instructions to an entire program.
Effectively, this approach bridges the gap between the granularity
of classic instruction level parallelism and thread or task level
parallelism. Another key point of this approach is that by using this
latency hiding approach, we can do with small simple cores with in-order
pipelines and smaller caches, allowing us to potentially put more
of these cores on one single chip. 

The Microgrid architecture is an implementation of the SVP concurrency
model \cite{SVP08}, which allows the expression of concurrency and
dependencies down to the level of loop level parallelism. One of the
contributions of the author to this research direction was the development
of a software run-time%
\footnote{From here on referenced as \emph{SVP-ptl}, the SVP POSIX Threads Library%
} that implements SVP on top of POSIX threads \cite{SVP-PTL09}, so
that we could test SVP programs on modern multi-core hardware. Of
course thread creation is 4 to 5 orders of magnitude slower, but it
did allow us to explore many aspects of SVP. Furthermore it has led
to research into a system level software SVP implementation allowing
finer grained scheduling with less overhead, and we are currently
working on applying this to the experimental 48-core Intel SCC \cite{IntelSCC10}
research chip. A future direction might be to consider a similar implementation
effort on a T3 system or future generation T-series, as well as other
future multi- or many-core architectures.

Coming from this angle, we see a lot of parallels between our Microgrid
architecture and the CMT approach of the T-series \cite{T3whitepaper2010}.
We were then first interested to investigate the boundaries of the
\emph{T3-4} system, and to see how the processor cores behave with
massively parallel workloads. As besides computationally bound, tasks
can either also be memory or I/O bound, we also try to measure the
maximum memory bandwidth. Unfortunately we did not have sufficient
I/O hardware available to do any sensible tests in the third direction.
As we are dealing with a system with this many cores, the mapping
of a task will have a great impact. We first investigate that in \prettyref{sec:Mapping-Experiment},
after which we measure the Integer arithmetic and Floating point arithmetic
throughput in \prettyref{sec:Computing-Throughput-Experiment} and
do a comparison with our reference machines. Then, we test the second
boundary of the machines with a Memory throughput test in \prettyref{sec:Memory-Throughput-Experiment}.
The last two test parts are based on tests relevant for the SVP-ptl
run-time \cite{SVP-PTL09}, one set of tests in \prettyref{sec:Pthread-experiments}
tries to measure the properties of the implementation of POSIX threads
(or pthreads in short) on our test machines, and one set of tests
in \prettyref{sec:SVP-ptl-related-experiments} running actual tests
using our SVP-ptl run-time.

\section{The effect of Mapping}\label{sec:Mapping-Experiment}

\subsection{Thread mapping experiences}

With as many as 512 virtual processors, it is interesting to see what
the effect of mapping your threads are, and how Solaris handles this
automatically. The first thing that we had to figure out was how the
processor IDs that can be passed to the \textbf{\emph{processor\_bind()}}
function are related to the actual processors, cores and thread slots
on the \emph{T3-4}. In our initial experiment we assumed a 'dumb'
mapping, where thread 0 is bound to processor ID 0, thread 1 to processor
ID 1, thread 511 to processor ID 511, thread 512 to processor ID 0
again, and so on. However, as expected, this does not give the optimal
performance, so we tried to discover how these IDs are actually distributed.
Using the \textbf{\emph{psrinfo -pv }}command it turned out that IDs
0-127 are on physical processor 0, IDs 128-255 on processor 1, etc.
Then, several tests with two threads on one physical processor revealed
that every sequence of 8 IDs correspond to the eight thread slots
of one single core. For example, virtual processor IDs 0-7 turned
out to map to the eight thread slots of core 0 on physical processor
0. However, these measurements also revealed that one thread on ID
0 and a second one on one of IDs 4 to 7 also yielded a performance
improvement for Integer Arithmetic. This can be explained by the fact
that each pipeline has two Integer ALUs, as found in the SPARC T3
datasheet \cite{SPARC-T3}, which apparently are shared among a group
of four threads. Using the information about mappings gathered by
these short experiments lead to the definition of our \emph{Round
Robin }mapping scheme where every four threads are divided amongst
the four physical processors, and then every sixteen threads on a
physical processor are divided over all sixteen cores. This means
that when mapping 64 threads, every physical core in the system gets
assigned exactly one thread in thread slot 0. Then the next group
of 64 threads that have to be mapped are divided similarly, but are
mapped to the slots corresponding to the second ALU on each core;
i.e. threads 0-63 are mapped to thread slot 0 on each core and threads
64-127 are mapped to thread slot 4 on each core. threads 128-255 then
map to thread slots 1 and 5 respectively, and so on until 512 threads
have been mapped and the system is completely full and mapping starts
again from slot 0, core 0 on physical processor 0. For independent
threads this should be the most optimal mapping, as the maximum of
computing resources is exposed. For threads communicating through
shared memory for example this is obviously not a straightforward
mapping strategy. As a third mapping method we decided to run our
experiment without calling the \textbf{\emph{processor\_bind() }}function,
and let the Solaris scheduler decide how to map our threads.

\subsection{Experiment setup}

In our first experiment, we compare the impact on computational throughput
using the three mapping methods described in the previous section.
In order to measure this, we create $n$ independent threads that
each loop over a small input dataset of 128 items and perform 8 arithmetic
operations using one input data and the result of the previous iteration.
The 8 arithmetic operations consist of two additions, two subtractions
and four multiplications. The calculation that is done is exactly
the same for both the integer arithmetic and floating point arithmetic
throughput tests. This loop is then executed one billion times, yielding
in $8\cdot10^{9}$ integer or floating point operations per thread,
excluding the loop overhead. These arithmetic operations are carefully
composed so that the calculation could not be simplified and rewritten
to less operations - or at least, that's what we originally thought.
After compiling this code with both GCC~\cite{GCC} (Version 4.5.1
with optimization flags: \textbf{-O3 -mtune=niagara2 -mpcu=niagara2
-mvis}) and the Sun C Compiler (version 5.11 with optimization flags:
\textbf{-m64 -fast -native}), it turned out that the code compiled
by the Sun C Compiler performed an amazing 33\% better than GCC. After
inspecting the assembly code generated by both compilers, it turned
out that the Sun C Compiler very aggressively managed to unroll the
computational loop, even though it had dynamic bounds passed through
a void{*} referenced struct through the \textbf{\emph{pthread\_create}}
call. After unrolling the loop, it could combine several operations
from the different iterations of the loop and yield only 6 instead
of 8 operations per iteration. Therefore, in order to calculate throughput,
we compensated our measurements to $6\cdot10^{9}$ operations per
thread for the Sun C Compiler, and $8\cdot10^{9}$ operations for
GCC compiled code. The mapping experiment was carried out using the
Sun C Compiler, and the comparison with GCC is further made in \ref{sec:Computing-Throughput-Experiment}.

In order to try to start each thread's computational part at the same
time, each thread first sets up the required memory, calls the \textbf{\emph{processor}}\textbf{\_}\textbf{\emph{bind()}}
function and acknowledges that it has started before waiting at a
barrier constructed with \textbf{\emph{pthread\_cond\_wait()}}. When
all threads have checked in, the barrier is released and each thread
retrieves and stores the time from \textbf{\emph{clock\_gettime()}}
using \textbf{\emph{CLOCK\_REALTIME}}. Then they enter the computational
loop explained earlier, and after completion call \textbf{\emph{clock\_gettime()}}
again to store the time after the computation. In order to calculate
our computing throughput, we iterate over the stored time values and
measure the time from the first thread to start its computation, until
the time at which the last one completed. Of course this is not completely
accurate as not all threads will start at exactly the same time, but
the computational part is proportionally large to compensate for this.
Actually, there does not seem to be any better way to do this.

\subsection{Results}

\begin{figure}
\begin{centering}
\includegraphics[width=1\textwidth]{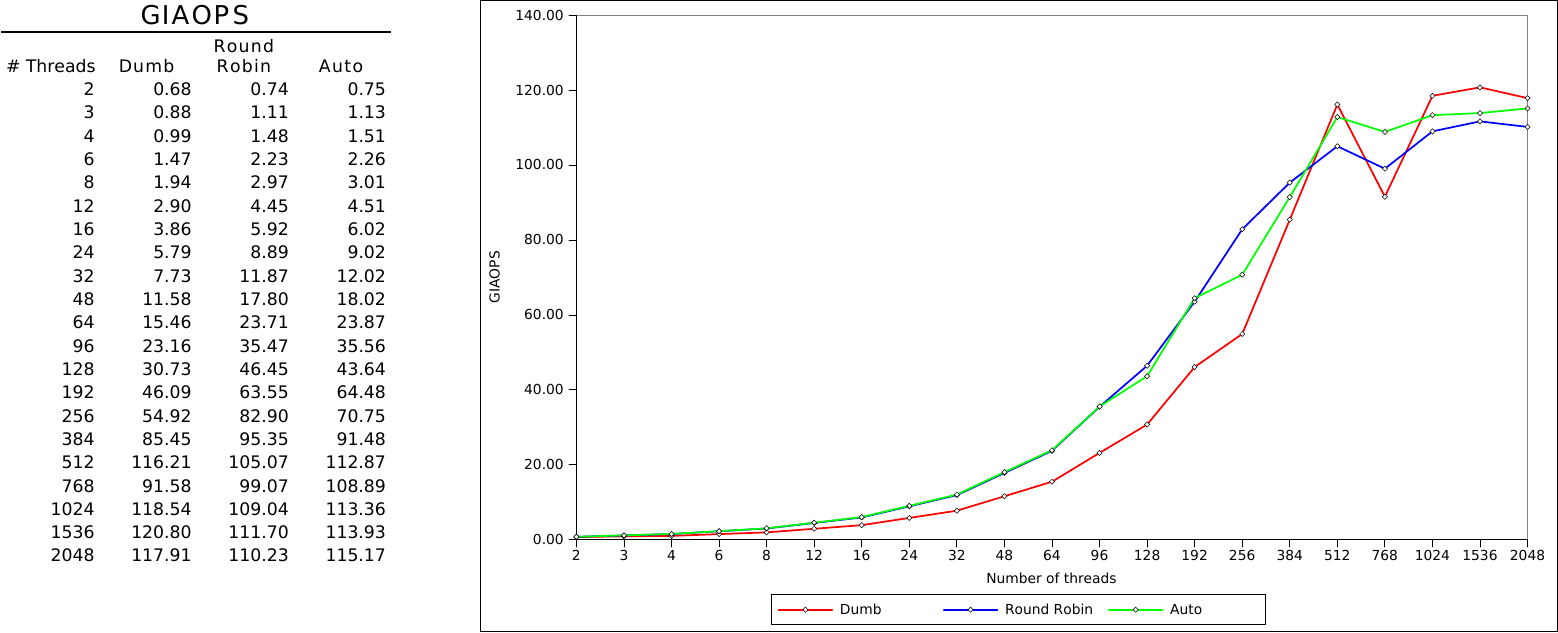}
\par\end{centering}

\caption{\label{fig:Mapping-Integer-Arithmetic-throughput}Integer Arithmetic
throughput on \emph{T3-4} with different mapping strategies}

\end{figure}
\begin{figure}
\begin{centering}
\includegraphics[width=1\textwidth]{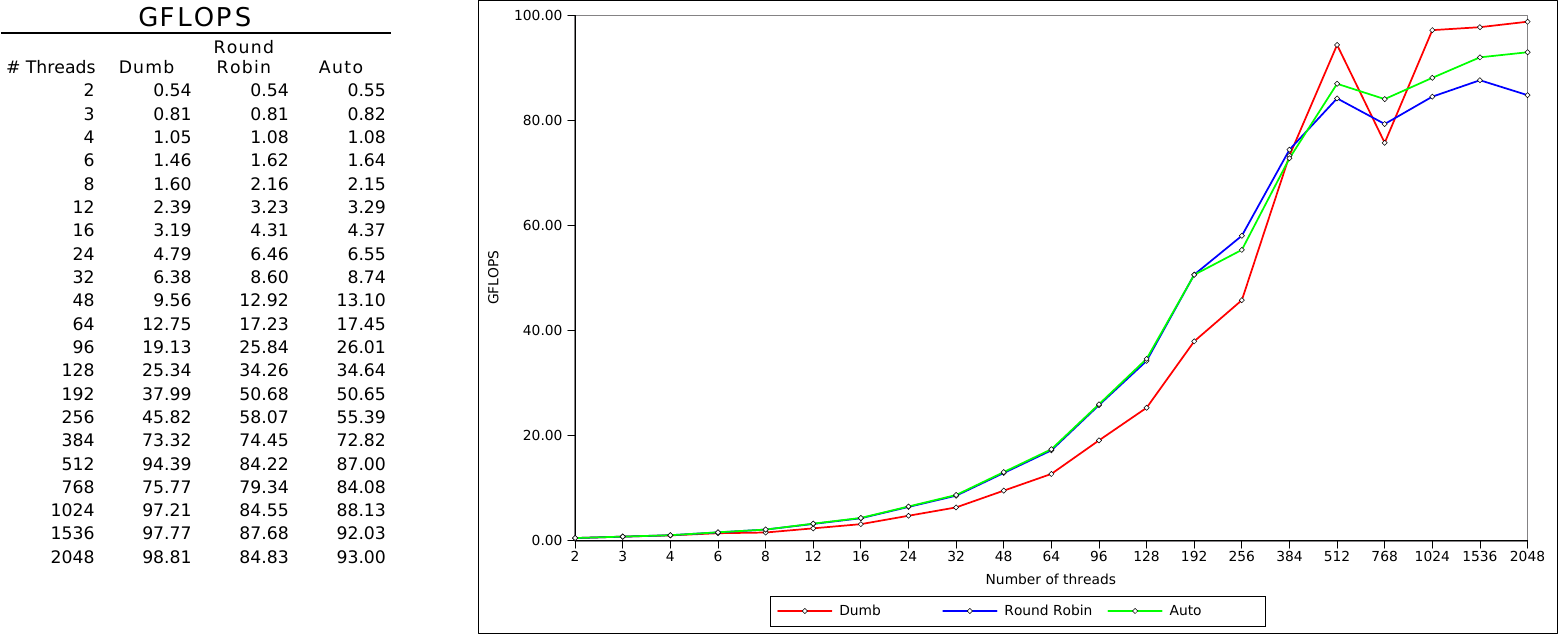}
\par\end{centering}

\caption{\label{fig:Mapping-Floating-point-Arithmetic-throughput}Floating-point
Arithmetic throughput on \emph{T3-4} with different mapping strategies}

\end{figure}
We have run the measurements using $2^{n}$ and $2^{n}+2^{n-1}$ threads
using the three different mapping strategies for both integer and
floating point performance. The result for Integers is shown in \prettyref{fig:Mapping-Integer-Arithmetic-throughput}
as GIAOPS%
\footnote{Giga-Integer Arithmetic Operations Per Second, i.e. $10^{9}s^{-1}$ %
}, we use this measure instead of, for example, MIPS as we are interested
in the throughput of instructions that perform the actual computation
of the kernel. As expected, the \emph{dumb} mapping performs worse
than the two other mappings due to load imbalances, until 512 threads
have been reached. The dip at 768 is caused by another load imbalance;
as physical processor 0 and 1 have twice the threads mapped to them
than processor 2 and 3. It is surprising that the throughput for multiples
of 512 threads performs slightly better than the other two mapping
strategies. We have no clear explanation for this, but possibly this
is a side effect of how the threads are started and synchronized with
the \textbf{\emph{pthread\_cond\_wait()}}\emph{ }barrier. As the threads
most likely reach the barrier in the order they have been created,
the OS probably puts them in a waiting queue in that order. As soon
as the barrier is released, this is the order that the threads are
woken up by the OS scheduler. When looking at the very fine grain
level, only one thread is woken up at a time, so after waking up thread
$n$ there needs to be some synchronization to wake up thread $n+1$.
With the \emph{dumb} mapping this synchronization is cheap as it is
almost always on the same physical processor, and even within the
same core. However, with the \emph{round robin} mapping, this synchronization
is \textbf{always} to another physical processor.

We see similar behavior in \prettyref{fig:Mapping-Floating-point-Arithmetic-throughput}
where we show the resulting performance for floating point calculations.
The main difference with the graph on integer performance is that
there is less difference between the \emph{round robin} and \emph{auto}
mapping at 128 and 256 threads. The difference for the integer arithmetic
performance is caused by the smart mapping to the two integer ALU
slots done by \emph{round robin}. As there is only one pipelined FPU,
this advantage is no longer present for the floating point experiment.
The rest of the graphs are similar to the integer experiment; \emph{round
robin} and \emph{auto} mappings perform equally well, except when
512 or more threads are mapped. The \emph{auto} mapping has the advantage
here that the OS scheduler is aware of other (system) tasks outside
the benchmark competing for CPU cycles, and can divide the resources
in a smarter, adaptive way.

\subsection{Conclusion}

The main conclusion of the mapping experiments is that the \emph{auto}
mapping where the mapping is left up to the Solaris scheduler performs
really well. Even though we could get a small advantage with \emph{round
robin }for certain cases\emph{, }overall the OS has a better idea
where and when to schedule the computations. Using this fact, we used
the \emph{auto} mapping for all our other experiments, unless other
mappings were relevantly interesting to investigate. 

Of course the \emph{auto }mapping will not always perform optimal;
for example, this is not necessarily the case when you have code that
communicates heavily through synchronization primitives or shared
memory. Then, placing your threads smartly on the system will potentially
be very beneficial. Our experiences also showed that it was not easy
to discover how the virtual processor IDs are laid out. We would recommend
to, for example, extend the \textbf{\emph{psrinfo -pv }}command to
provide more information on how virtual processor IDs are related
to actual cores. This is relevant, as we have shown the difference
between mapping to thread slots and mapping to cores by the difference
in performance for the \emph{dumb} and \emph{round robin} mappings.

As a secondary conclusion, not specific to the \emph{T3-4}, we were
really impressed by the Sun C Compiler 5.11; the fact that it very
aggressively managed to unroll and simplify an all but trivial loop
in our benchmark code was surprising, to say the least. Our comparison
compiler, GCC 4.5.1, did not manage to apply this optimization.

\section{Computing Throughput
Scalability}\label{sec:Computing-Throughput-Experiment}

In this experiment we try to measure the maximum computing throughput
the \emph{T3-4} can achieve, and compare it with the \emph{X4470}
and \emph{Magny Cours} systems.

\subsection{Experiment setup}

The experiment setup is the same as for the mapping experiment in
\prettyref{sec:Mapping-Experiment}, using the \emph{auto} mapping
strategy. It was performed on the \emph{T3-4} with two versions; one
compiled with the Sun C Compiler and one with GCC using the versions
and flags mentioned in the previous section, and using the same compensation
for the number of arithmetic instructions. We then compiled the code
for the \emph{X4470} using GCC 4.1.2 with \textbf{-mtune=core2 -march=core2
-mfpmath=sse,387 -msse -msse2 -msse3 -msse4a}, and for the \emph{Magny
Cours} using also GCC 4.1.2, with \textbf{-mtune=amdfam10 -march=amdfam10
-mfpmath=sse,387}. We did not apply mapping techniques to the two
reference machines, so we leave this up to the Linux scheduler. As
we were interested in the maximum throughput we could achieve, each
measurement was carried out three times and the maximum value was
used for the result.

\subsection{Results}

\begin{figure}
\begin{centering}
\includegraphics[width=1\textwidth]{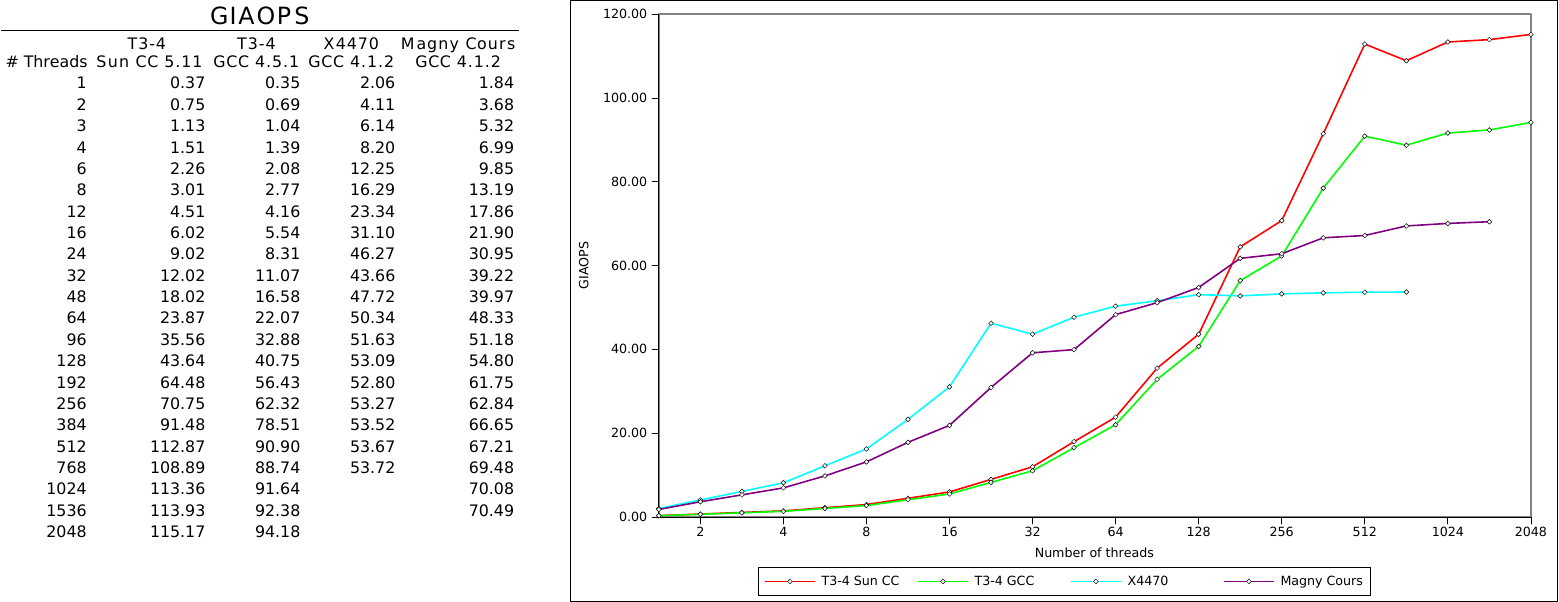}
\par\end{centering}

\caption{\label{fig:Integer-Arithmetic-Throughput}Integer Arithmetic Throughput
Scalability}

\end{figure}
\begin{figure}
\begin{centering}
\includegraphics[width=1\textwidth]{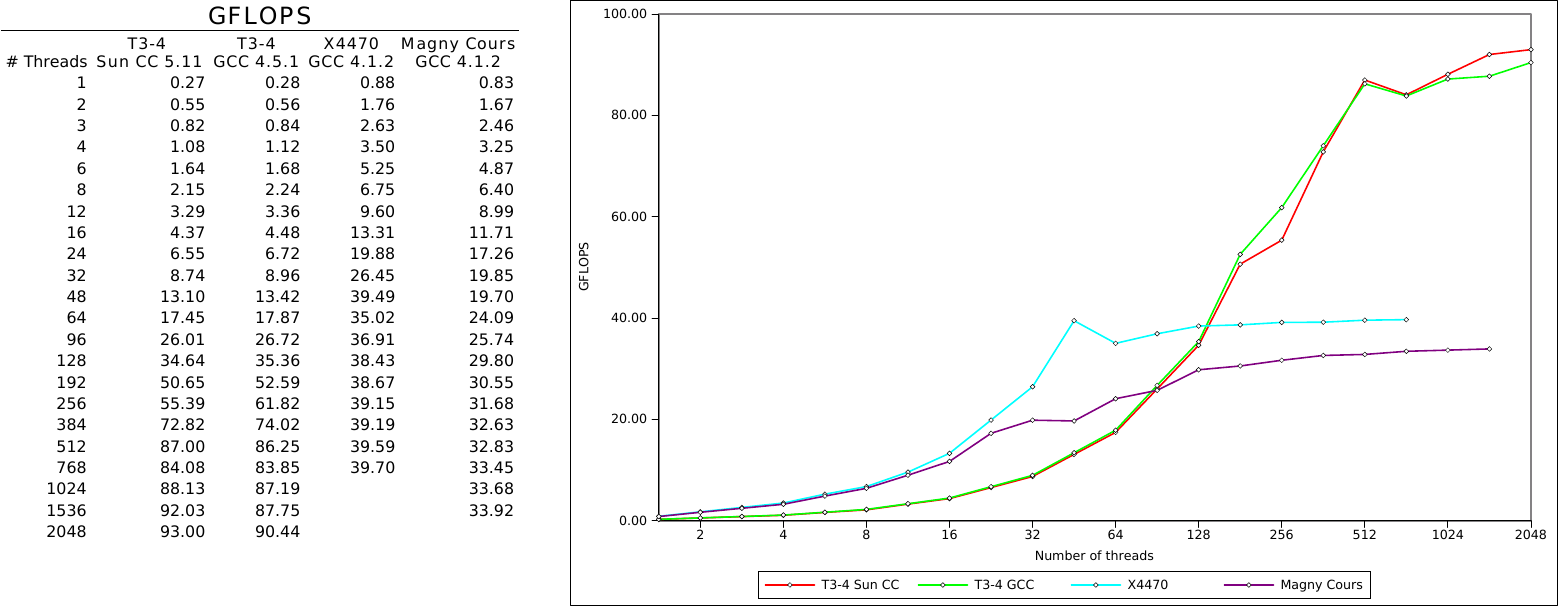}
\par\end{centering}

\caption{\label{fig:Floating-point-Arithmetic-Throughput}Floating-point Arithmetic
Throughput Scalability}

\end{figure}
The result of measuring integer arithmetic throughput on all the three
systems is shown in \prettyref{fig:Integer-Arithmetic-Throughput}.
The first thing that stands out is that the code generated by the
Sun C Compiler has a much higher integer arithmetic throughput on
the \emph{T3-4} than the code generated by GCC. This is related to
the loop optimization discussed earlier, as the GCC generated code
has much more instructions in between to handle the loop, which compete
for integer ALU resources in the pipeline. This is confirmed by looking
at the floating point throughput in \prettyref{fig:Floating-point-Arithmetic-Throughput},
where we do not observe such a large difference. A second thing that
we can observe about the \emph{T3-4}s performance is that it really
scales up to 512 threads, requiring all thread slots on all cores
to be in use to maximally utilize the ALUs and FPUs in the system.

Comparing to the \emph{X4470} and \emph{Magny Cours}, it is no surprise
that these surpass the \emph{T3-4} in single thread arithmetic throughput,
as they have highly complex super scalar out of order execution pipelines,
as well as a higher clockrate. The Xeon cores of the \emph{X4470}
outperform the Opterons in the \emph{Magny Cours}, even though they
run at a slightly slower clock. The scalability of the \emph{X4470}
is as expected, the throughput gradually increases until all 24 cores
are filled up, and it gets a small boost at 48 threads due to its
Hyperthreading after which it flattens out. This is even more the
case for the floating point throughput, where the Hyperthreading really
helps to get the maximum out of the FPUs. The scaling of the \emph{Magny
Cours} system is a bit strange, as it still keeps scaling after 48
threads, almost doubling the throughput eventually at 1024 threads.
We have the impression that this could be either caused by the Linux
scheduler having difficulty distributing work over such a high number
of processors, or because of the input dataset for all threads which
is stored in one contiguous block of memory. We observe the same effect
for both integer and floating point throughput.

An interesting note can be made on comparing both integer and floating
point throughput between the \emph{T3-4} and the two reference systems.
When we look at the throughput of a single thread, not surprisingly,
is outperformed by the two other systems due to its more simple in
order pipeline. However, when we compare the throughput of a single
core, i.e. in the \emph{T3-4} filled with 8 threads, we have to look
again at the result table for the \emph{dumb} mapping in \prettyref{fig:Mapping-Integer-Arithmetic-throughput}
and \prettyref{fig:Mapping-Floating-point-Arithmetic-throughput}.
Here we see that one single core of the \emph{T3-4} has a throughput
of 1.98 GIAOPS, which is very comparable to the 2.06 and 1.84 of the
reference systems, but even surpasses them with 1.60 GFLOPS against
0.88 and 0.83 respectively per core. The 1.60 GFLOPS per core show
that the fully pipelined FPUs can really accept an operation every
cycle as the cores clock at 1.65 GHz. Actually this last comparison
did not take the Hyperthreading into account on the \emph{X4470 }as
we did not measure this separately, but judging from the scaling in
\prettyref{fig:Floating-point-Arithmetic-Throughput} this would come
down to roughly 1.7 GFLOPS. Also we would like to note that it is
unlikely that the any of the systems could use vector instructions
(i.e. MMX/SSE or VIS) to speed up the calculation, and a separate
benchmark would have to be constructed to compare the performance
of these extensions.

\subsection{Conclusion}

It really takes 8 threads per core on the \emph{T3-4} to saturate
all its execution units, and the system really scales up to 512 thread
workloads computationally wise. The performance of a single core (not
a single thread) is very comparable to that of the \emph{X4470} and
\emph{Magny Cours} system, given it has enough threads to execute.
This exposes the nice property of how the \emph{T3-4}s cores do latency
hiding between threads, very similar to the Microgrid architecture
designed by the CSA group, as discussed in \prettyref{sec:Background}.
Even though we did not test the vector instruction extensions on any
of the test systems, we can still conclude that the \emph{T3-4} is
likely to deliver a large computational throughput for embarrassingly
parallel scientific computing applications.

\section{Memory Throughput Scalability}\label{sec:Memory-Throughput-Experiment}

After we explored the computational bounds of the three systems in
the last section, we will now turn to the effect of memory throughput.
As all three systems are 4-socket based with 4-channel DDR3-SDRAM
memory interfaces per socket, this should be an interesting comparison.

\subsection{Experiment Setup}

The test is set up similarly as the previous two experiments, in which
all threads were released (as good as) simultaneously with a barrier,
and the time is measured between the first starting thread and the
last thread to complete. Before the threads enter the barrier, but
after (if applicable) they have assigned themselves to a virtual processor,
each allocates a memory block of 256 MB. Besides allocating it with
\textbf{\emph{malloc()}}, each thread also writes to the complete
block with \textbf{\emph{memset()}}, to prevent the operating systems
from {}``optimizing'' the allocation using pagefault handlers and
only actually allocating once a page is accessed. For the writing
experiment, after the barrier and reading out the timer, the thread
writes zeros to the entire 256 MB block using a \textbf{\emph{bzero()}}
call. We picked the memory block sufficiently large to have one long
consecutive write that would be large enough for a clean measurement.
Also not repeatedly reading/writing to the same memory ranges avoids
any possible caching behavior as we really wanted to measure the throughput
that we can obtain from the off-chip memory. For the read experiment
we treat the 256 MB block as an array of integers (that has been zeroed
before the barrier), loading each element from memory by adding it
to a register within a loop.

As each physical processor has four memory channels associated, we
again investigate the influence of mapping threads with different
strategies on the \emph{T3-4}. Similarly as in \prettyref{sec:Mapping-Experiment},
we use the \emph{dumb}, \emph{round robin}, and \emph{auto} mapping
schemes again. We do not specify a mapping on the reference machines,
leaving up to the Linux scheduler where to map the threads.

\subsection{Results}

\begin{figure}
\begin{centering}
\includegraphics[width=1\textwidth]{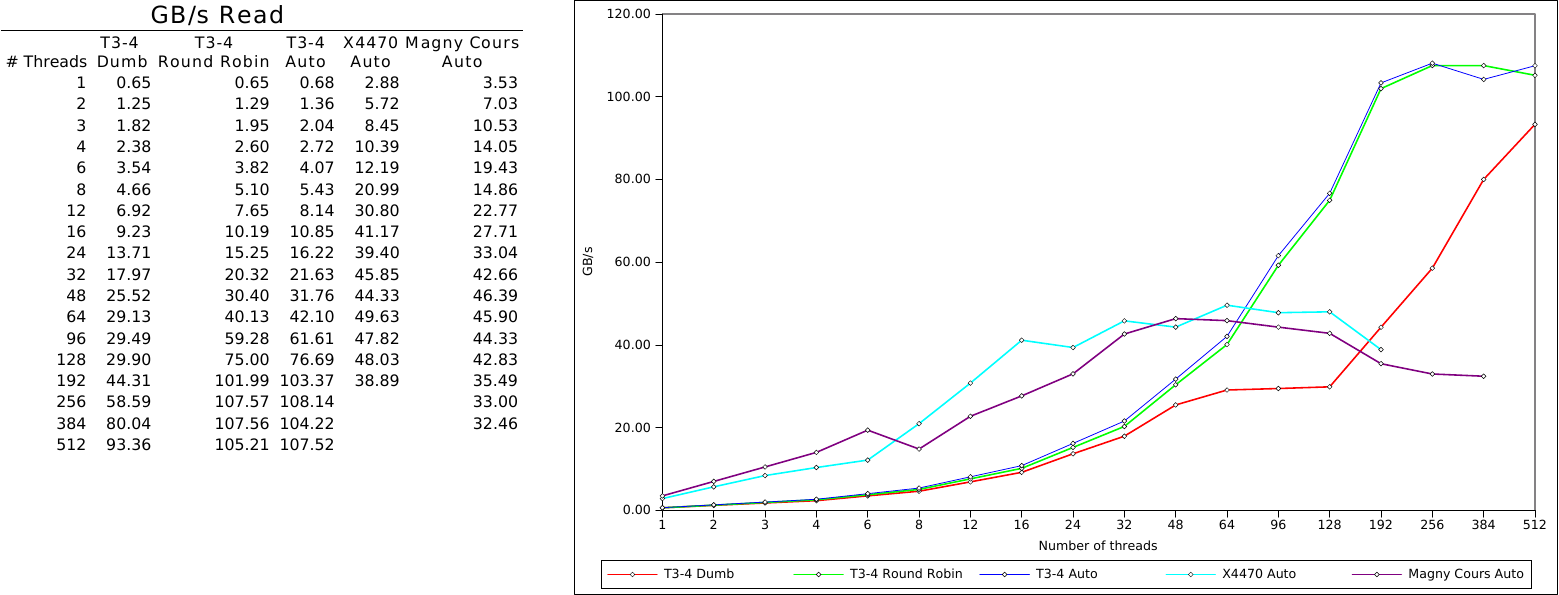}
\par\end{centering}

\caption{\label{fig:Memory-Read-Throughput}Memory Read Throughput}

\end{figure}
\begin{figure}
\begin{centering}
\includegraphics[width=1\textwidth]{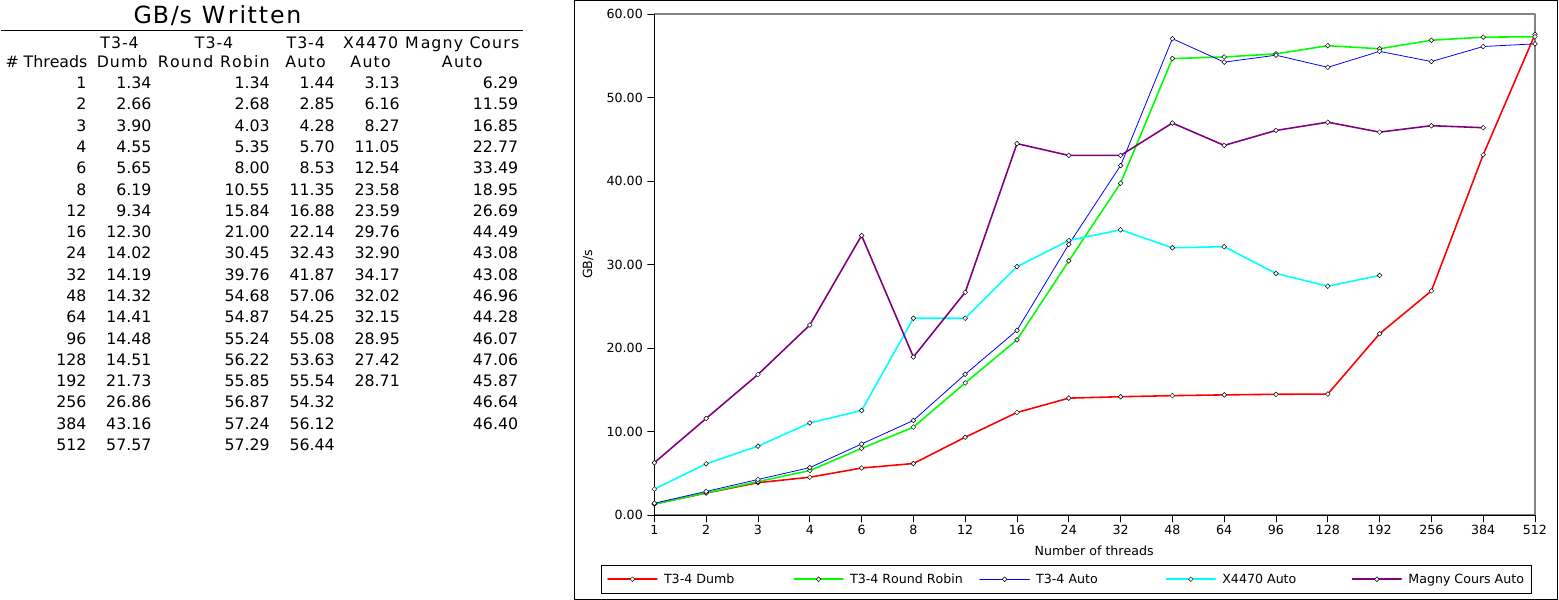}
\par\end{centering}

\caption{\label{fig:Memory-Write-Throughput}Memory Write Throughput}

\end{figure}
The results for reading throughput are shown in \prettyref{fig:Memory-Read-Throughput}
and for writing throughput in \prettyref{fig:Memory-Write-Throughput}.
From the \emph{dumb} mapping in the reading throughput graphs we can
see from 64 to 128 threads that we saturate the throughput which one
physical processor can achieve at 29 GB/s. And for the \emph{round
robin} and \emph{auto }mapping there is little difference once again,
and at 256 threads they manage to saturate the bandwidth of all four
physical processors, coming to an aggregate of 108 GB/s. As for writing,
the \emph{T3-4} saturates a single physical processor using the \emph{dumb}
mapping at 24 threads with 14 GB/s throughput, and all four at 57
GB/s using 48 threads. The reason that the achieved throughput is
lower for writing than for reading can be explained in two ways. When
reading large blocks from memory, prefetching techniques will pay
off, cache lines are filled, and after the first read the next few
will hit in the data cache. Secondly, for memory writes, the cache
line is probably fetched first on an initial write miss and then at
some point written back to memory again effectively having to use
the memory bandwidth twice, and potentially, snooping messages have
to be sent out to every other physical processor on every write.

The results for the \emph{X4470 }and \emph{Magny Cours} machines are
also shown, and again, for single or few threads they outperform the
\emph{T3-4} system in achieved memory bandwidth. For writing, the
\emph{Magny Cours} has a big drop in throughput around 8 and 12 threads,
but this might be caused by Linux not mapping the threads efficiently.
At 48 threads the \emph{Magny Cours }unsurprisingly reaches the peak
of its throughput performance, but remarkably it has a better writing
than reading throughput; something we repeatedly measured, and that
we currently can not explain. Possibly a side effect of the coherency
scheme they use. The \emph{X4470} performs as expected, quite similar
in reading behavior as the \emph{Magny Cours}, with already at 16
threads almost saturating memory read and write throughput. It is
interesting that for both reading and writing, a jump in performance
is made when going from 6 to 8 threads. It could be the case that
Linux by default maps the 6 threads to one physical processor as it
has 6 cores, and that at 8 threads we see a larger increase in performance
as threads are moved to other physical processors and use the independent
memory controller there.

\subsection{Conclusion}

The \emph{T3-4}s memory bandwidth scales up well for a high number
of threads, and above the point that it saturates, it does not suffer
much penalty. It delivers twice the throughput for reading than our
reference systems. This throughput combined with the dataflow like
scheduling of the threads in the cores provides enough bandwidth to
accommodate many threads which due to latency tolerate can provide
a large computing throughput as seen in \prettyref{sec:Computing-Throughput-Experiment}.
This property is what the CSA group also achieves in their Microgrid
architecture, as discussed in \prettyref{sec:Background}, and mentioned
before in the conclusion of \prettyref{sec:Computing-Throughput-Experiment}.

\section{Experiments with POSIX threads}\label{sec:Pthread-experiments}

\begin{wrapfigure}[11]{o}{0.33\textwidth}%
\begin{centering}
\includegraphics[width=0.25\textwidth]{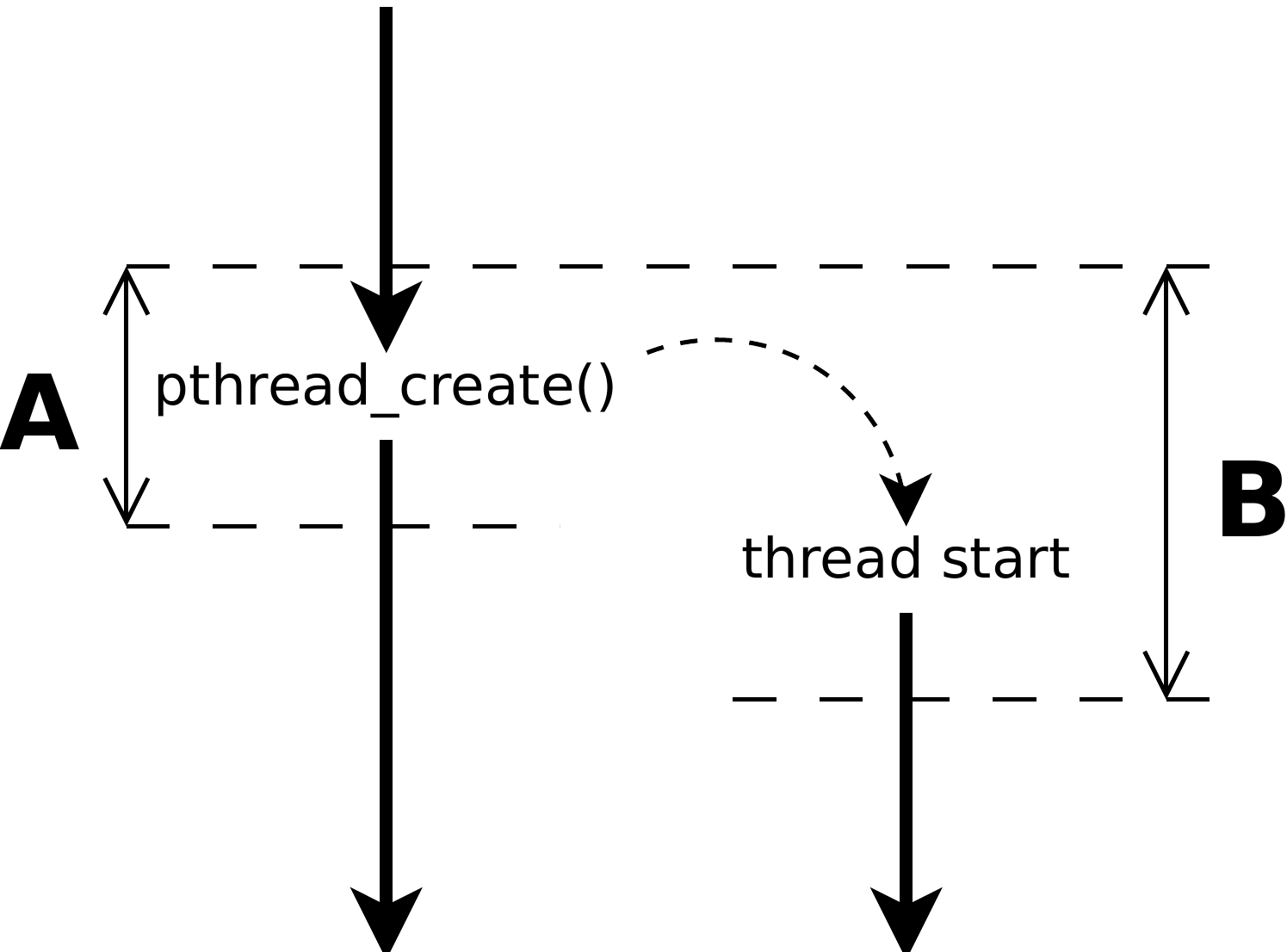}
\end{centering}

\caption{\label{fig:thread-create-diagram}Measured intervals for thread creation}
\end{wrapfigure}%
As we now have identified the throughput bounds of the computational
and memory side in the previous sections, we now will look at how
the implementation of pthreads behaves on the three tested systems.
We normalize each measured interval using the clockspeed of each system,
in order to come to an estimated number of cycles that the measured
mechanism takes. First we will look into the costs of thread creation,
then thread synchronization with mutexes, and finally thread synchronization
with conditionals.

\subsection{Thread Creation}

\begin{figure}[t]
\begin{centering}
\includegraphics[width=0.53\textwidth]{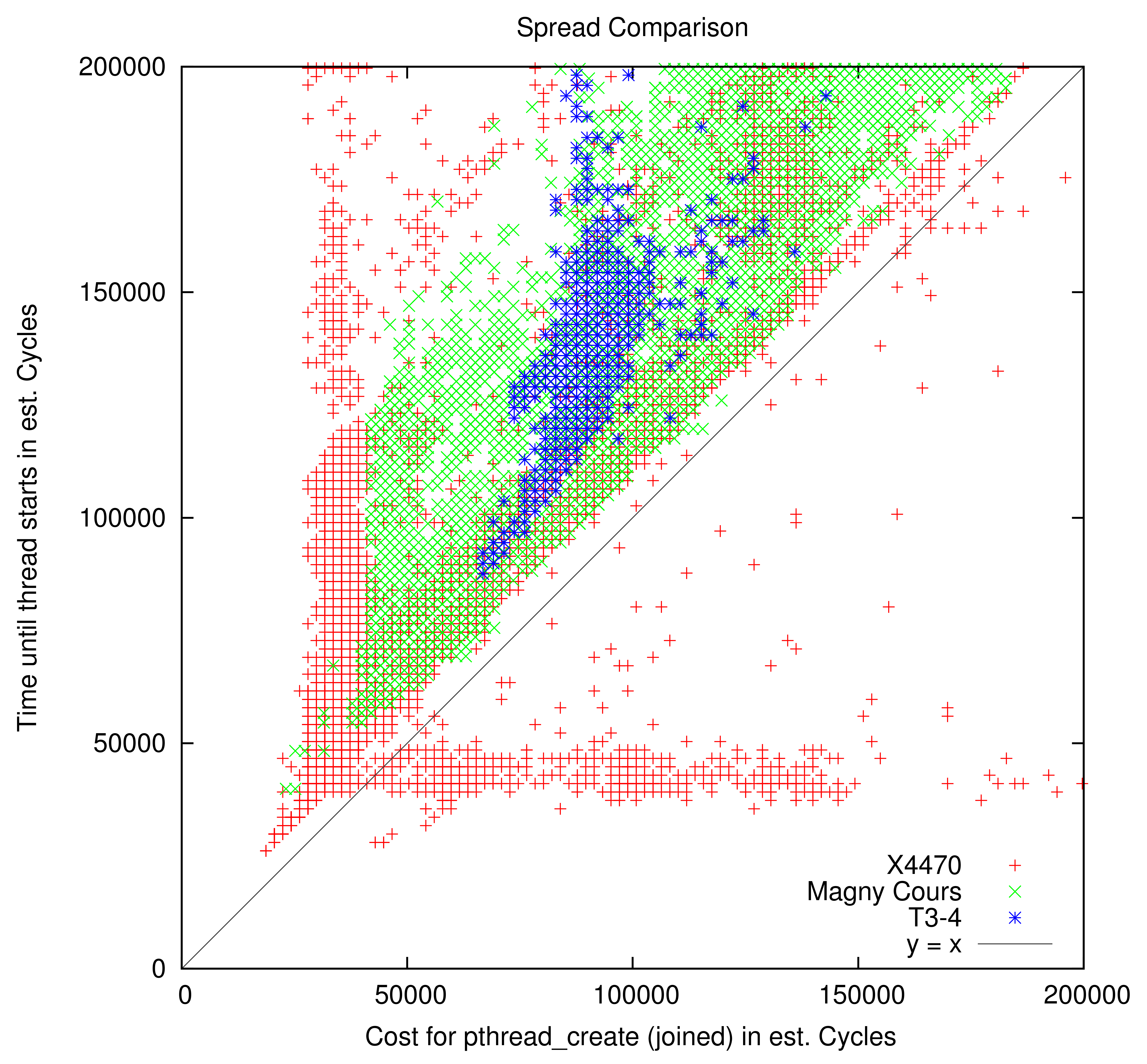}\includegraphics[clip,width=0.47\textwidth]{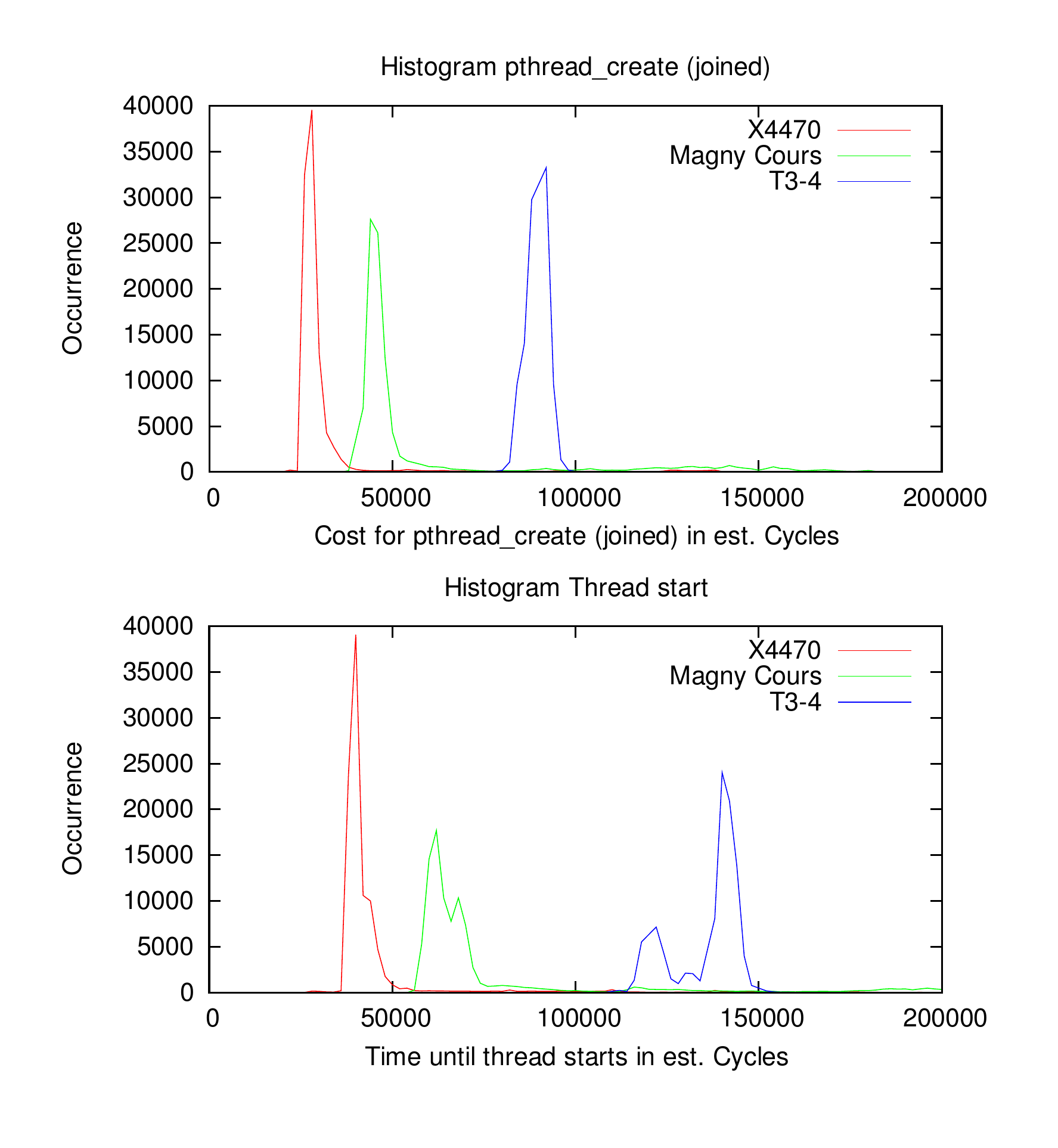}
\par\end{centering}

\begin{centering}
\includegraphics[width=1\textwidth]{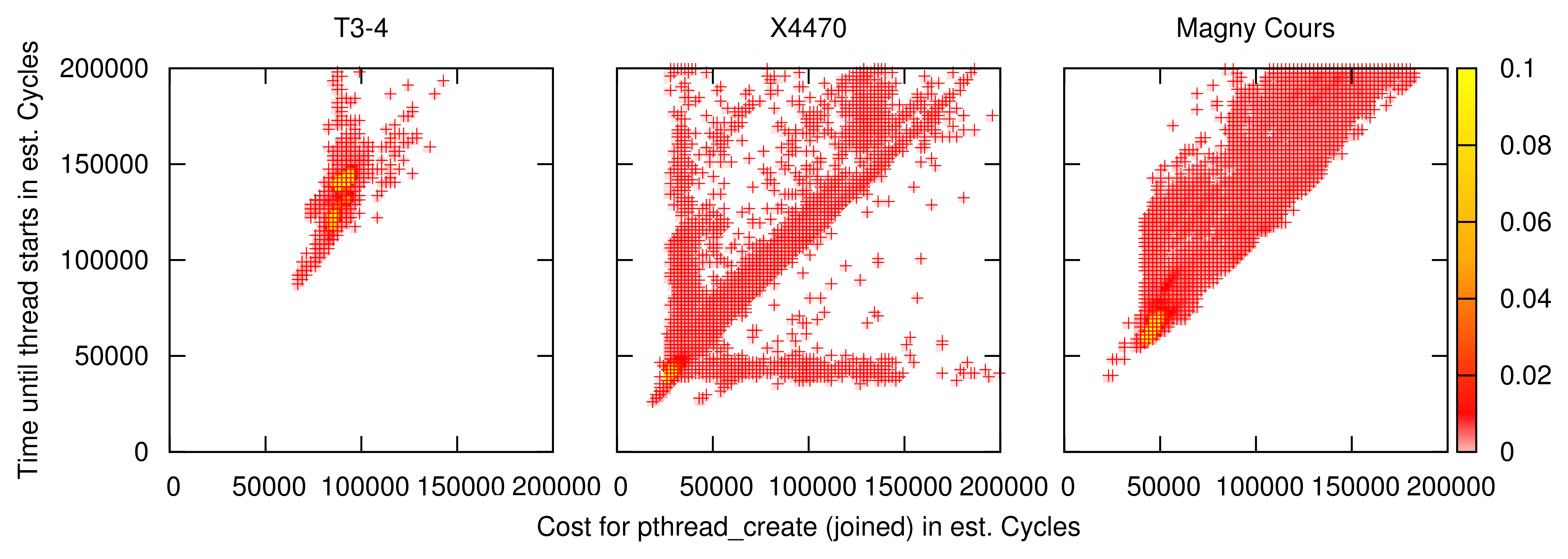}
\par\end{centering}

\caption{\label{fig:pthread-create-join}Measured behavior of \textbf{\emph{pthread\_create()}}
with joinable threads}

\end{figure}
In our first experiment with pthreads we will look at the overhead
for starting threads. When using pthreads, a thread can be created
in two modes; joinable or detached. Threads that are joinable allow
another thread that has their identifier (usually the parent) to wait
for its termination to receive back a return code, while detached
threads will simply disappear on termination. For these experiments
we have measured the times for the \textbf{\emph{pthread\_create()}}
call to return, and the time from the \textbf{\emph{pthread\_create()}}
call until the created thread starts executing its thread function.
A diagram representing these two times is shown in \prettyref{fig:thread-create-diagram}.

We have measured the overhead for thread creation using $10^{5}$
samples on all our three systems with joinable threads, and the results
are shown in \prettyref{fig:pthread-create-join}. The figure at the
left-top is a scatter plot showing the relation between time \emph{A}
and time \emph{B} as shown in \prettyref{fig:thread-create-diagram},
overlaying the results for all three machines. In this plot also a
line $y=x$ is added, as all points below this line indicate situations
in which the created thread started before the \textbf{\emph{pthread\_create()}}
call returned in the parent. As this plot does only show how the measurements
are spread, but not show their intensity, the three plots at the bottom
and two histograms on the side are generated. The three plots at the
bottom show an intensity plot for the dataset of each system, where
we can see clearly where the common case is located. The histograms
give an impression how the values \emph{A} and \emph{B} are overall
spread, regardless of their relation.

\begin{figure}[t]
\begin{centering}
\includegraphics[width=0.53\textwidth]{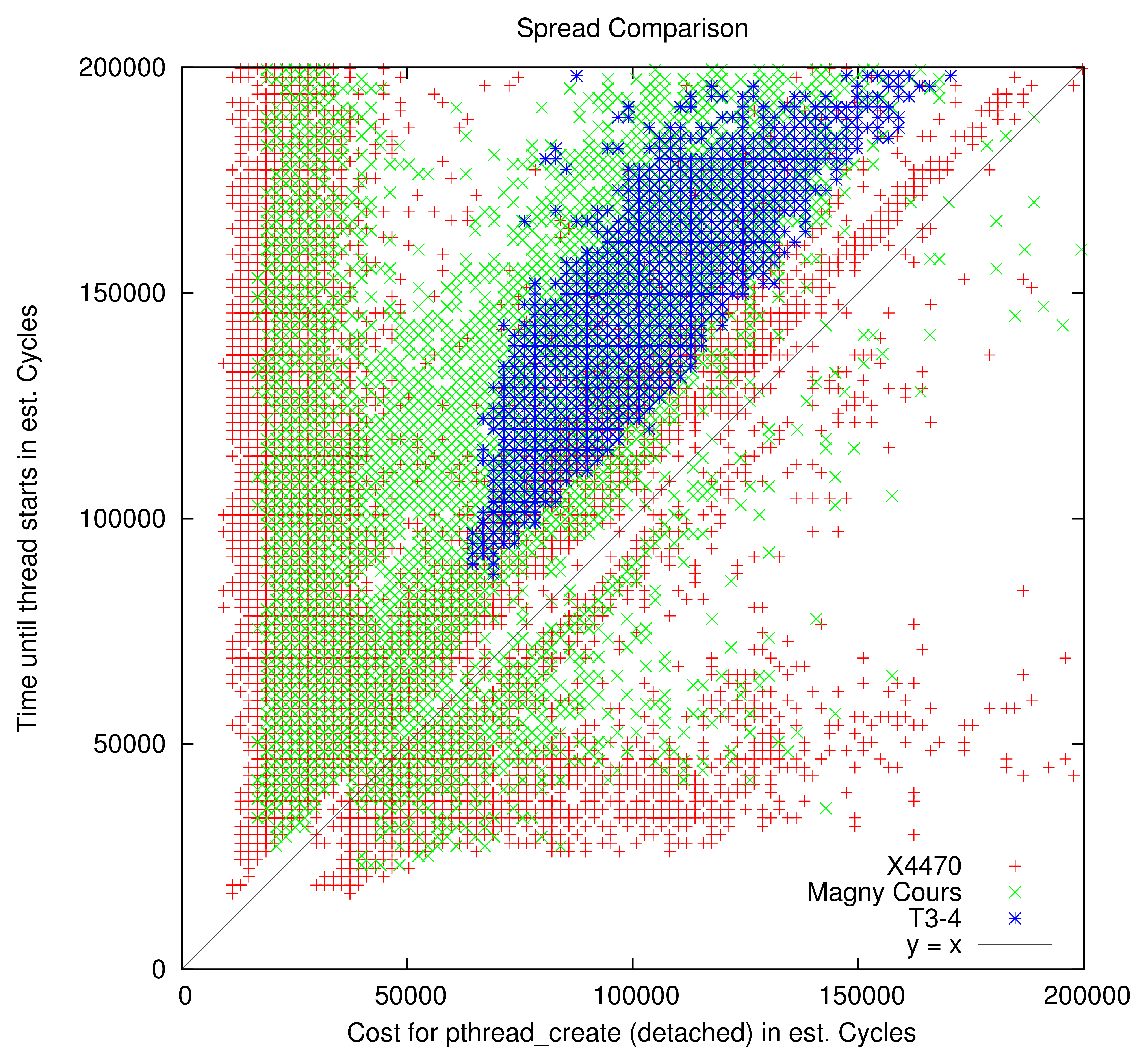}\includegraphics[clip,width=0.47\textwidth]{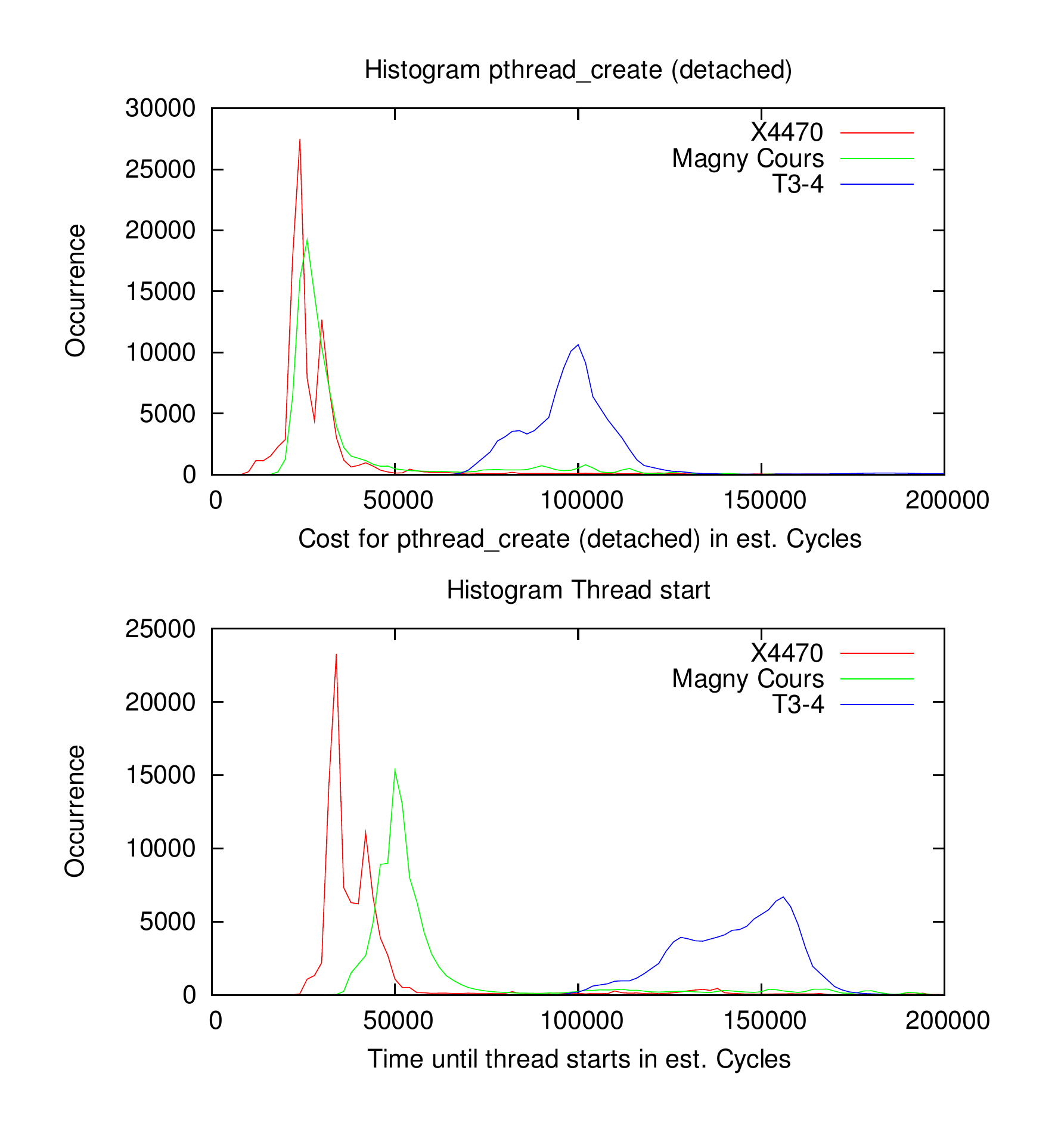}
\par\end{centering}

\begin{centering}
\includegraphics[width=1\textwidth]{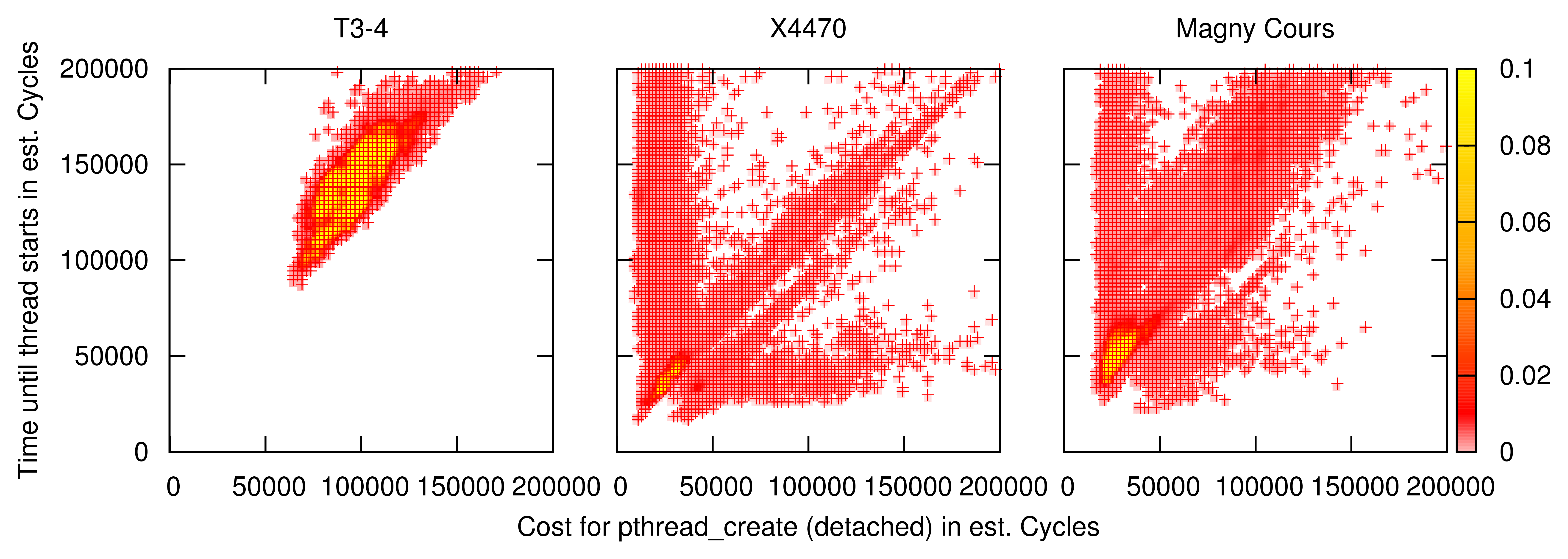}
\par\end{centering}

\caption{\label{fig:pthread-create-detach}Measured behavior of \textbf{\emph{pthread\_create()}}
with detached threads}

\end{figure}
The first thing we notice from the data presented in \prettyref{fig:pthread-create-join}
is that the Intel Xeon based \emph{X4470} system has the quickest
thread creation, with 80\% of the threads starting within 38-44 KCycles
after the create call. This is much more widespread on the AMD Opteron
based \emph{Magny Cours} system where 72\% of the threads are started
within 58-70 KCycles. However, if we look at the \emph{T3-4}, 70\%
of the threads are started in 138-146 KCycles. It is not a surprise
that thread creation is slow on the \emph{T3-4}, but this is not specific
to the \emph{T3}-\emph{4, }as we already observed this on an UltraSPARC
machine running Solaris 8 some time ago in \cite{SVP-PTL09}, and
also on other tests on a single CPU UltraSPARC running Solaris 10.
Judging from the positioning of the points representing the \emph{T3-4}
in the plot, it seems that most of these cycles are spent in the parent
thread, as it takes equally long to execute the \textbf{\emph{pthread\_create()}}
call. This is apparently not always the case for the \emph{X4470}
system, so probably the scheduler decides to place the newly created
thread on the same core, switches context, and only completes the
\textbf{\emph{pthread\_create()}} call later on. It is interesting
that this behavior is not seen on the \emph{Magny Cours} system, as
both systems are running the Linux 2.6.18 kernel. Probably the scheduler
makes different decisions depending on the architecture of the system,
or perhaps it exploits Hyperthreading by creating the new thread in
the second hardware thread slot on the \emph{X4470}.

We have repeated the same experiment as discussed above, but now for
threads that are created in detached mode. The results of these measurements
are presented in a similar way, in \prettyref{fig:pthread-create-detach}.
The first thing we notice is that the cost for creating a thread has
reduced considerably on the Linux systems. Furthermore, the \emph{Magny
Cours} platform now also shows threads that start before the \textbf{\emph{pthread\_create()}}
call returned, and on average the call returns earlier. For Solaris
on the \emph{T3-4}, the creation of threads is still as expensive,
but there is more variation in the cost. On the \emph{X4470}, 74\%
of the threads take 32-44 KCycles to start, 69\% of the threads in
44-58 KCycles on the \emph{Magny Cours}, and 77\% of the threads in
124-160 KCycles on the \emph{T3-4}.

\subsection{Mutex Synchronization}

\begin{wrapfigure}[11]{o}{0.33\textwidth}%
\begin{centering}
\includegraphics[width=0.25\textwidth]{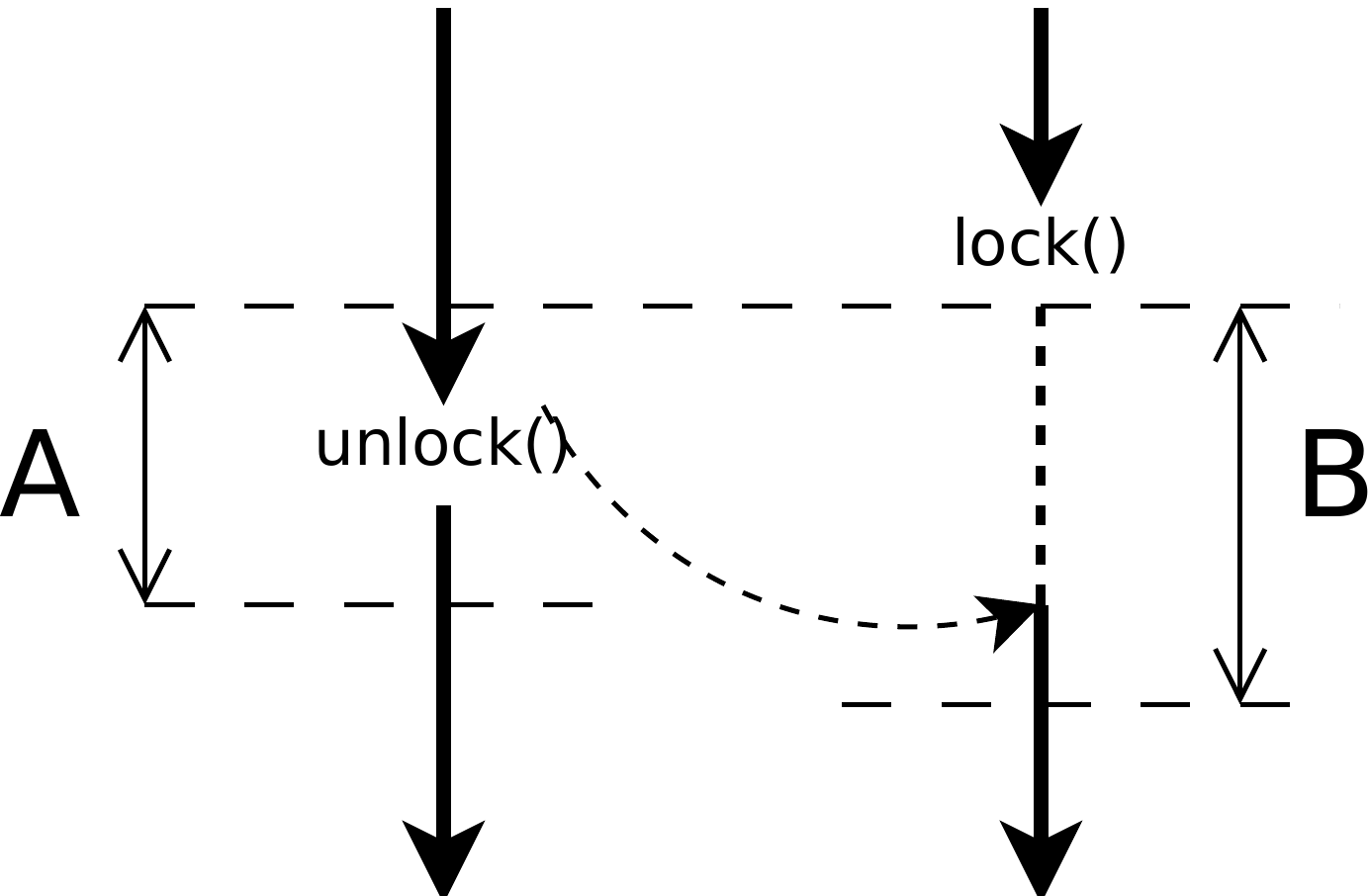}
\end{centering}

\caption{\label{fig:mutex-diagram}Measured intervals for contended mutex}

\end{wrapfigure}%
Most modern operating systems implement an optimization to handle
non-contended mutexes fully in userspace, to avoid the cost of making
a system call and context switch to kernel mode. Linux for example
implements this with a \emph{futex}, a fast userspace mutex based
on atomic operations provided in the instruction set. We measured
the overhead of locking and unlocking a mutex a thousand times by
a single thread, and calculated the average cost of the combination
of these two operations. On the \emph{T3-4}, this takes 432 cycles
most of the time, against 46 cycles on the \emph{X4470}, and 56 cycles
on the \emph{Magny Cours} test system. We then developed a test to
see how quick a thread wakes up that was waiting on a mutex. We measure
the time from the start of the unlock operation until the thread that
was waiting was scheduled, as shown in \prettyref{fig:mutex-diagram}.

\begin{figure}
\begin{centering}
\includegraphics[width=0.53\textwidth]{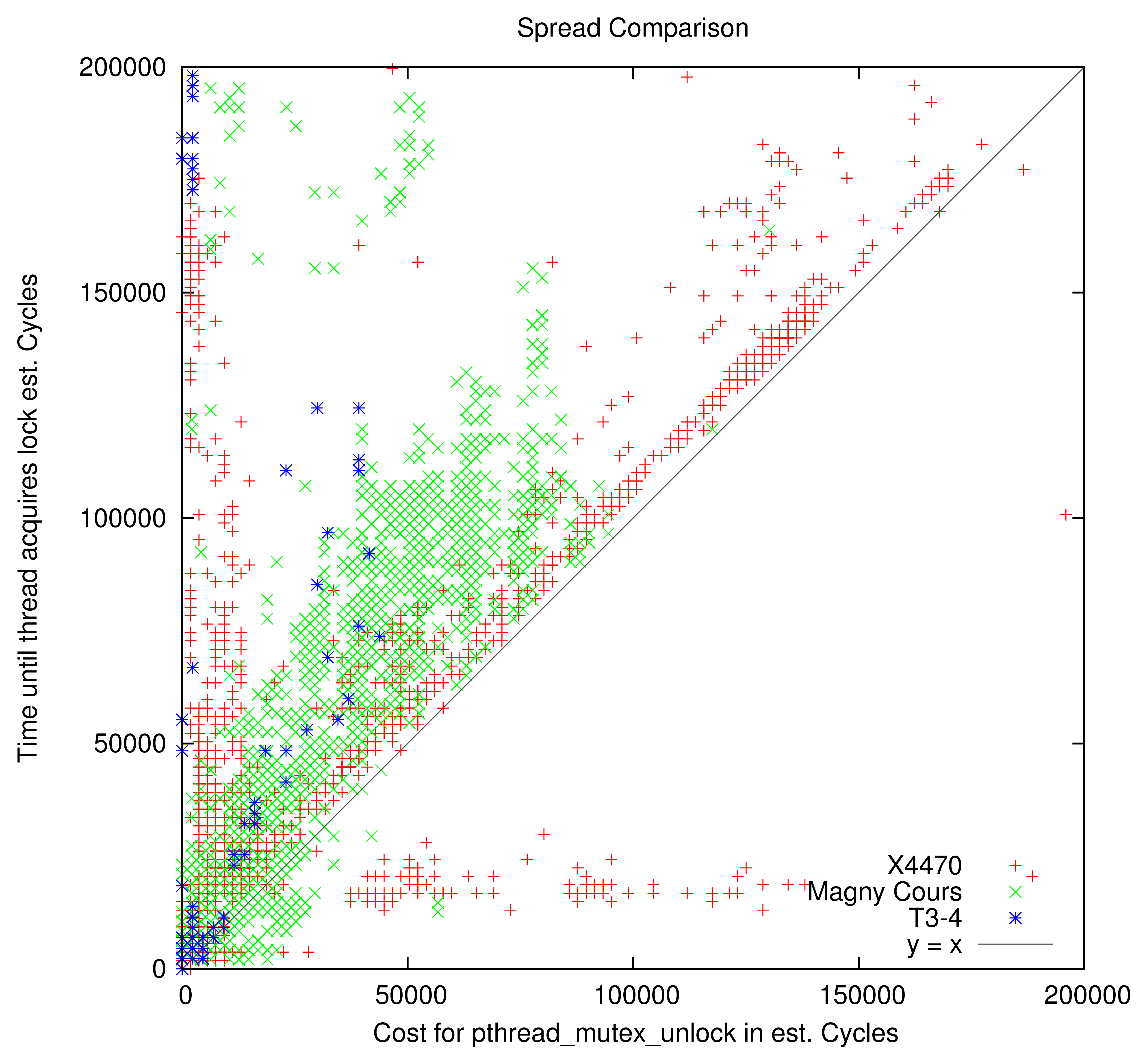}\includegraphics[clip,width=0.47\textwidth]{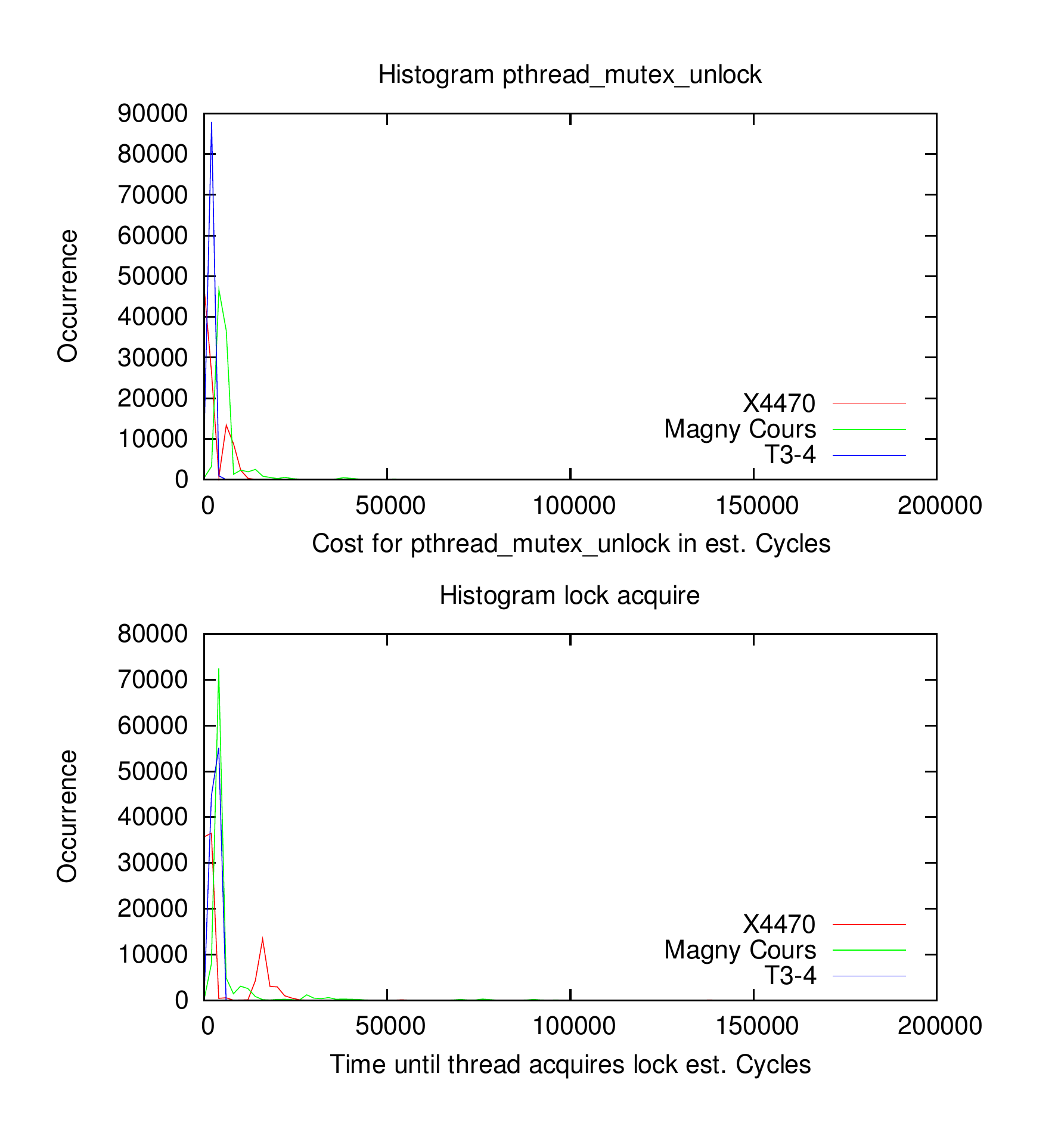}
\par\end{centering}

\begin{centering}
\includegraphics[width=1\textwidth]{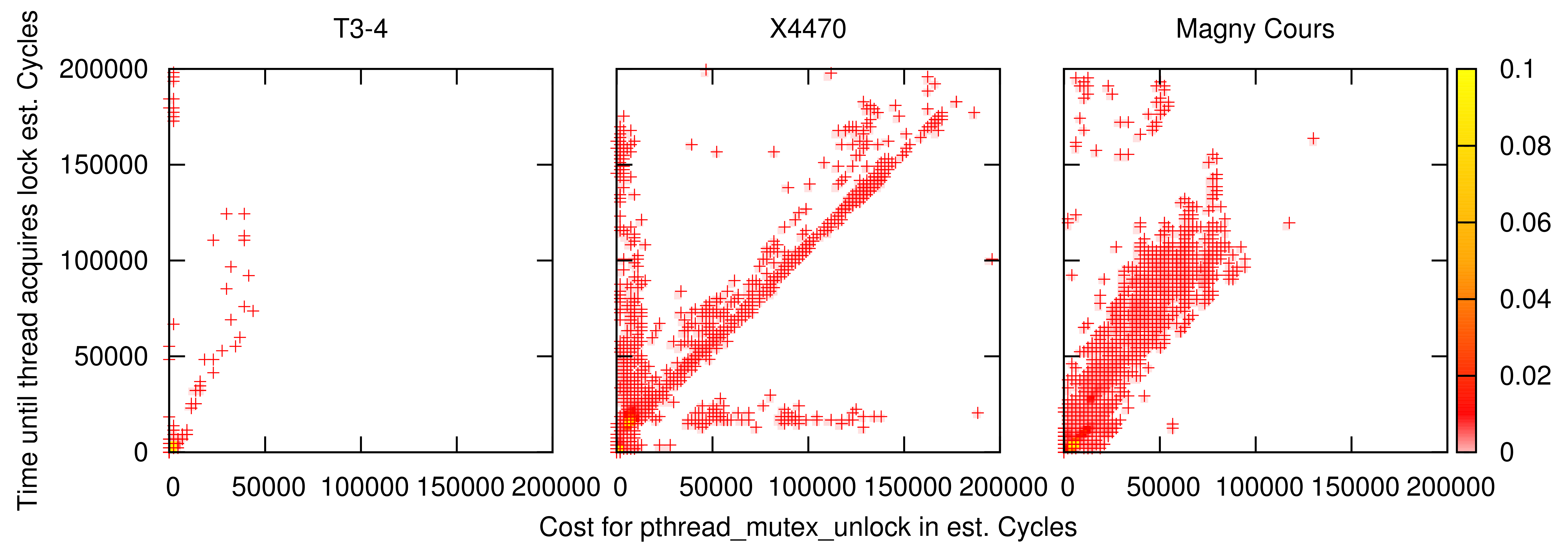}
\par\end{centering}

\caption{\label{fig:mutex-behavior}Measured behavior of \textbf{\emph{pthread\_mutex\_unlock()}}}

\end{figure}
The results of $10^{5}$ synchronizations between two threads through
a mutex are shown in \prettyref{fig:mutex-behavior}, with the same
style of diagrams as in the previous section. There was no explicit
mapping for these two threads, so their (relative) placement is unknown.
For the \emph{T3-4}, 99\% of the threads start in less than 4 KCycles
after the mutex is released, while the unlock call takes around 2-4
KCycles 99\% of the cases. These figures are less concentrated for
the \emph{X4470} and the \emph{Magny Cours}; for the \emph{X4470},
only 73\% of the threads started in less than 4 KCycles, (95\% in
less then 10 KCycles), and 72\% of the calls to unlock took around
2 KCycles, and 90\% in only less than 18 KCycles. For the \emph{Magny
Cours} system, 86\% of the threads are started in 4-6 KCycles, 80\%
of the unlocks took 2-4 KCycles, and 92\% of the unlocks were done
in less than 14 KCycles.

\subsection{Conditional Synchronization}

\begin{wrapfigure}[11]{o}{0.33\textwidth}%
\begin{centering}
\includegraphics[width=0.25\textwidth]{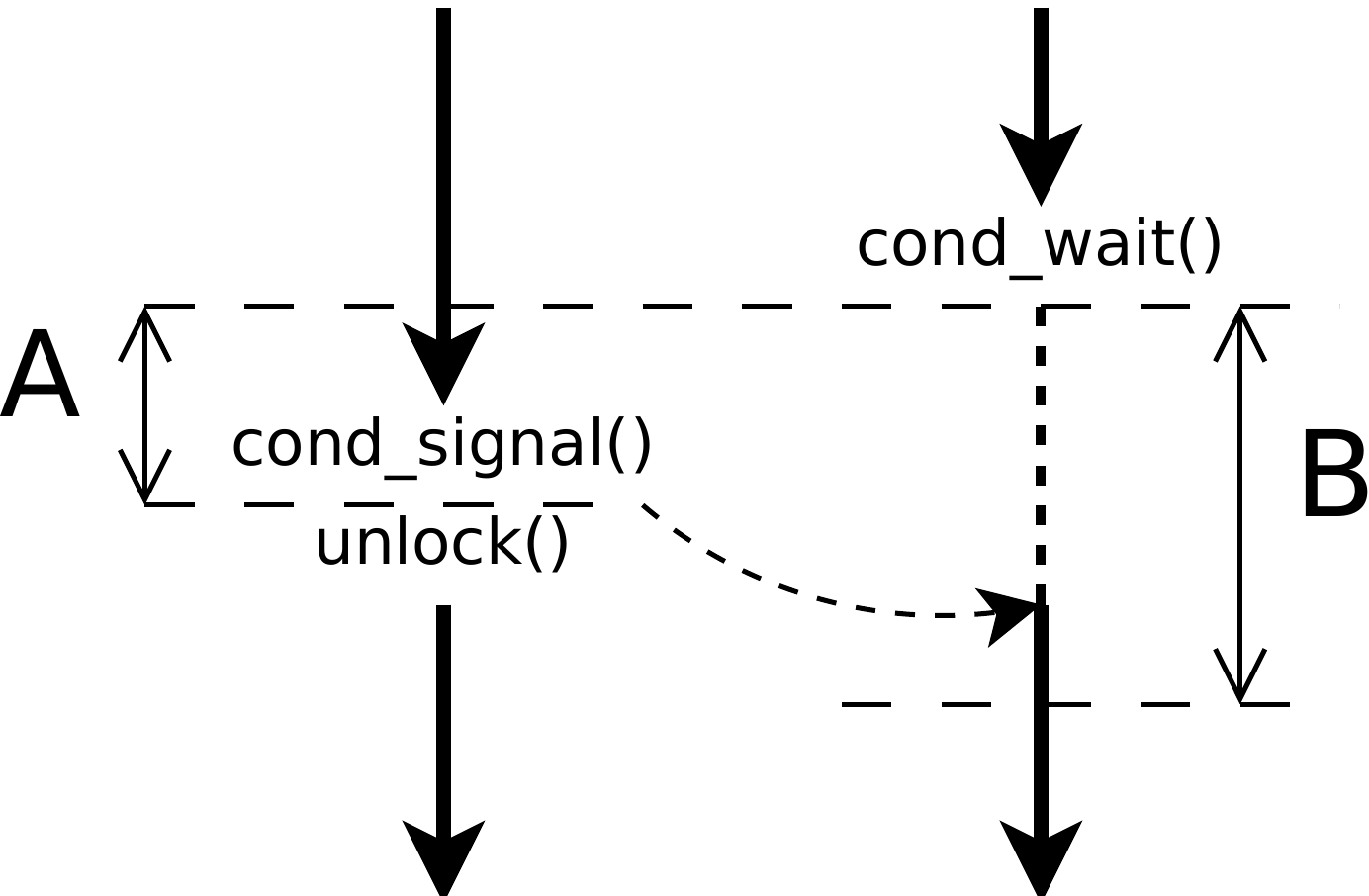}
\end{centering}

\caption{\label{fig:Diagram-conditional}Measured intervals for conditional
synchronization}
\end{wrapfigure}%
A second common way of synchronization between two or more threads
is through a conditional and a value guarded by a mutex. A conditional
can be signaled in two ways; a single signal which will only wake
up one thread that is suspended on the conditional, or a broadcast
that will wake up all threads. We measure the latency of both methods
between two threads, with the times we measured depicted in \prettyref{fig:Diagram-conditional}.

\begin{figure}
\begin{centering}
\includegraphics[width=0.53\textwidth]{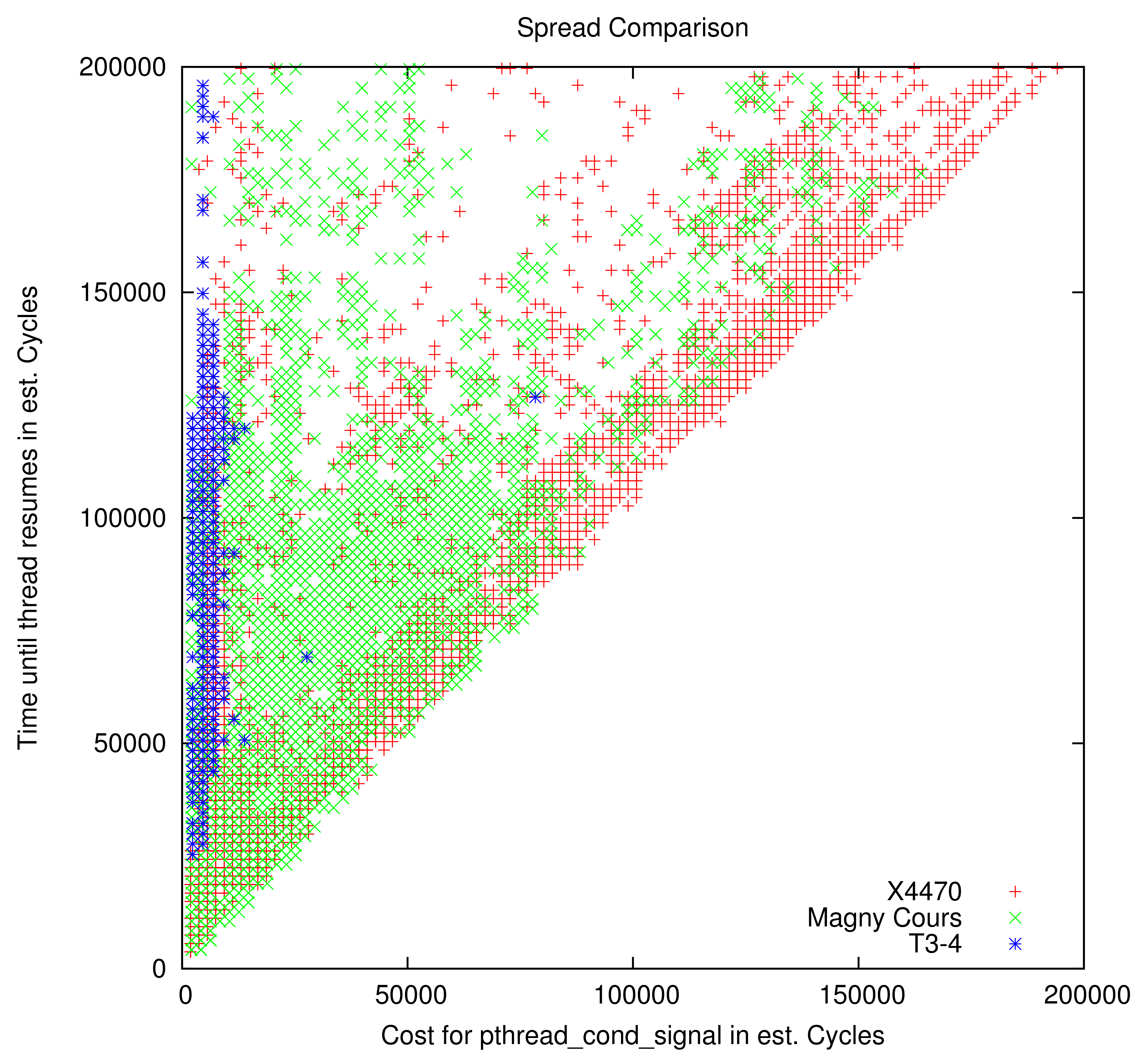}\includegraphics[width=0.47\textwidth]{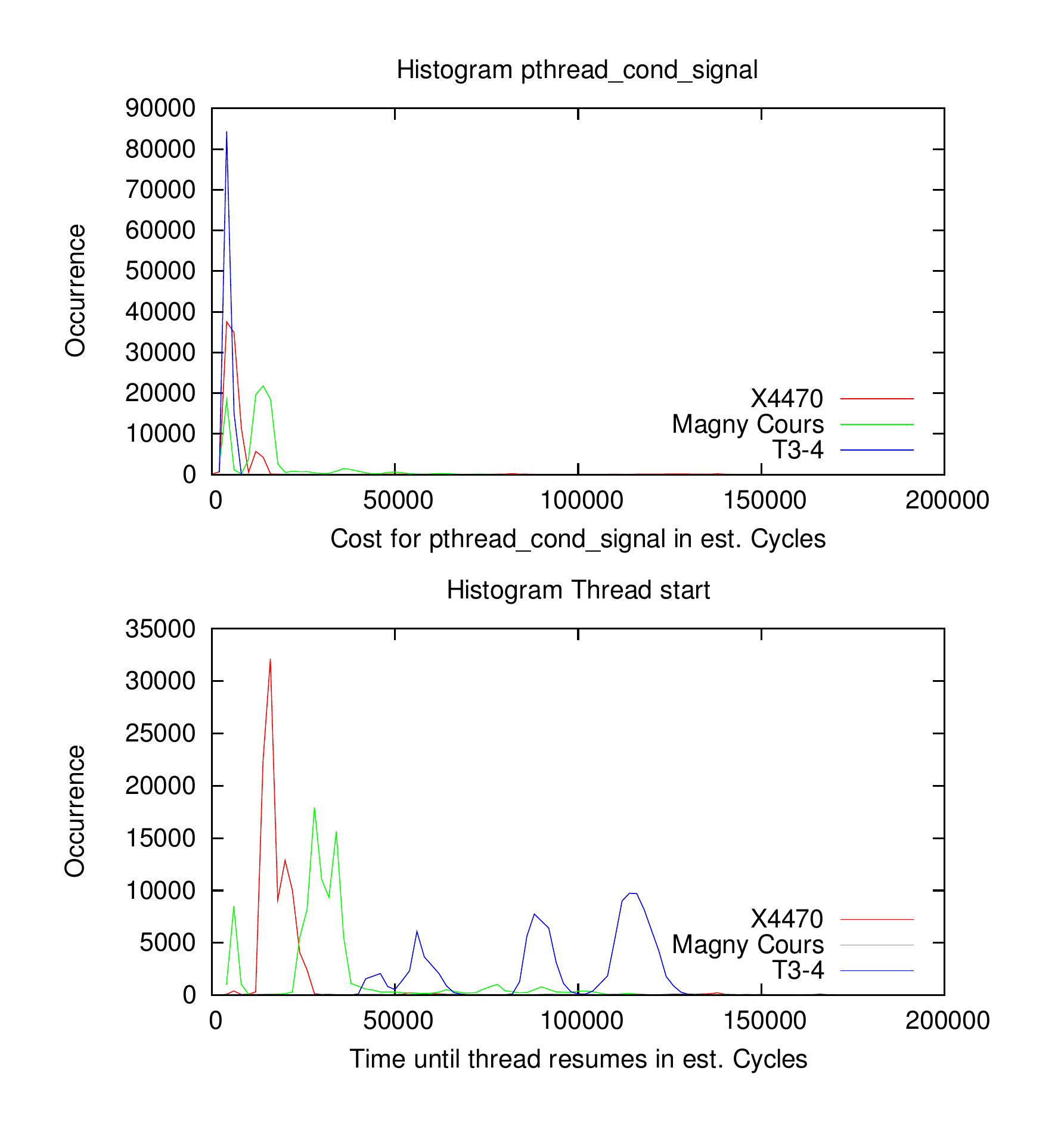}
\par\end{centering}

\begin{centering}
\includegraphics[width=1\textwidth]{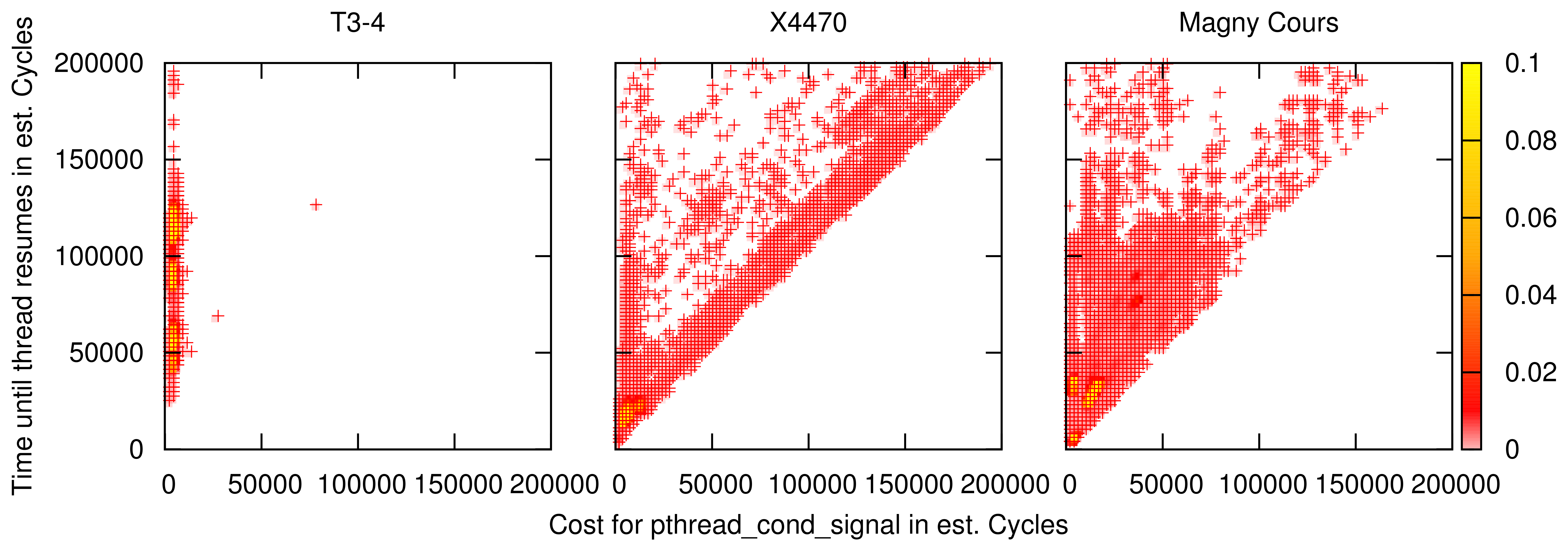}
\par\end{centering}

\caption{\label{fig:cond-signal-behavior}Measured behavior of \textbf{\emph{pthread\_cond\_signal()}}}

\end{figure}
The latency of sending a \textbf{\emph{pthread\_cond\_signal()}} until
the target thread wakes up are shown in \prettyref{fig:cond-signal-behavior}
for $10^{5}$ measurements, again using the similar visualizations
of spread, intensity and relation. It is clear that the \emph{T3-4}
system has a predictable time in signaling the conditional with 99.9\%
of the calls taking around 2-6 KCycles, while on the \emph{X4470}
83\% of the calls completes after 4-8 KCycles and on the \emph{Magny
Cours}, 18\% of the calls completes around 4 KCycles, but another
62\% takes around 12-18 KCycles. It is interesting to see how the
signaled threads respond to this in relation, as this is quite spread
for the \emph{T3-4}, on which the wake-up times are spread around
three peaks; 18\% is around the 42-62 KCycles area, 25\% around 86-94
KCycles, and then 49\% around 110-124 KCycles. The \emph{X4470} is
quite consistent as 86\% of the threads wake up after 14-22 KCycles.
On the \emph{Magny Cours} we also observe something interesting; 10.5\%
of the threads wake up after 4-8 KCycles, corresponding with a part
of the 18\% of the signal calls that completed on that system in 4
KCycles, the second peak of 73\% of the threads wake up at 24-36 KCycles.

\begin{figure}
\begin{centering}
\includegraphics[width=0.53\textwidth]{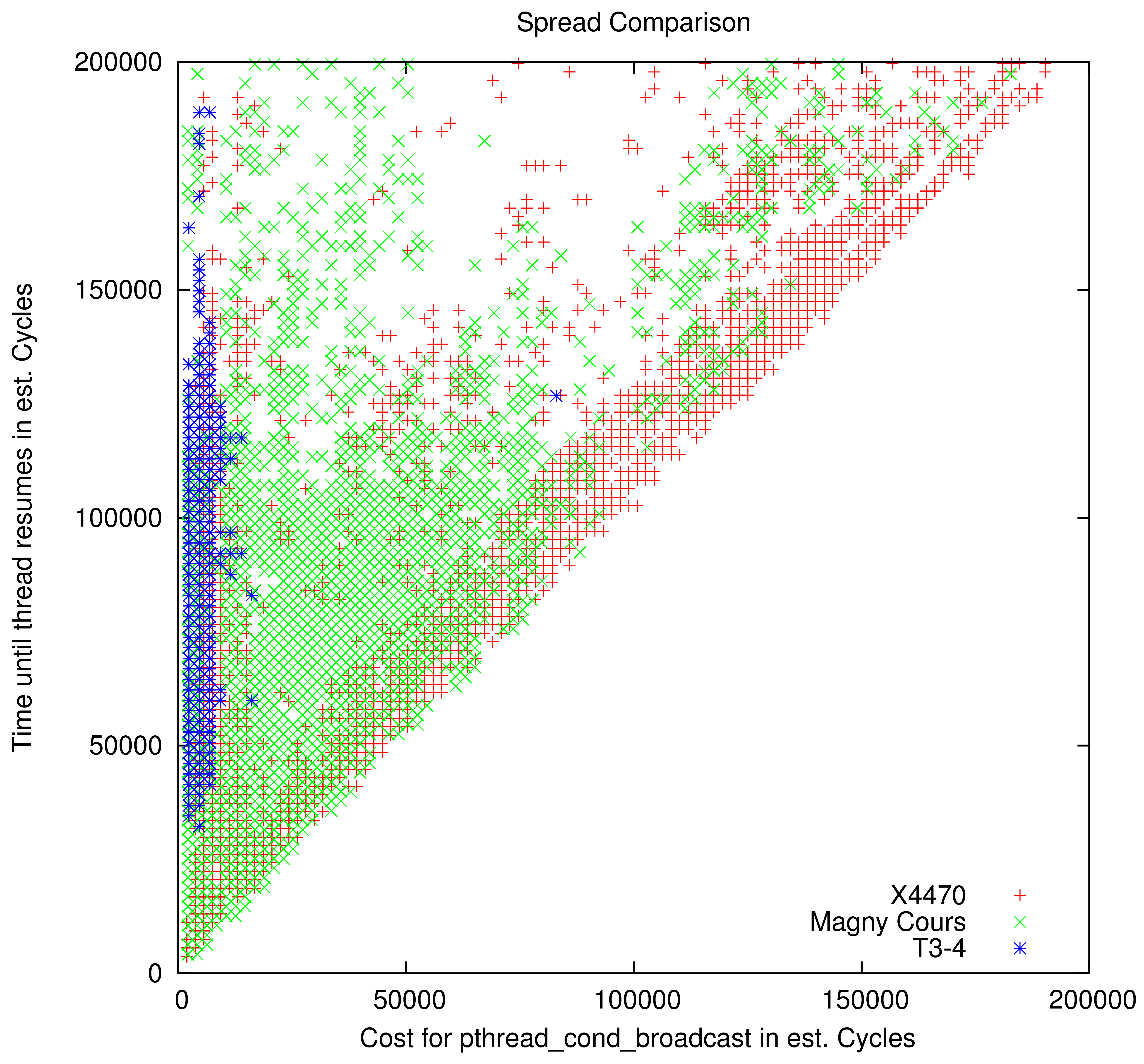}\includegraphics[clip,width=0.47\textwidth]{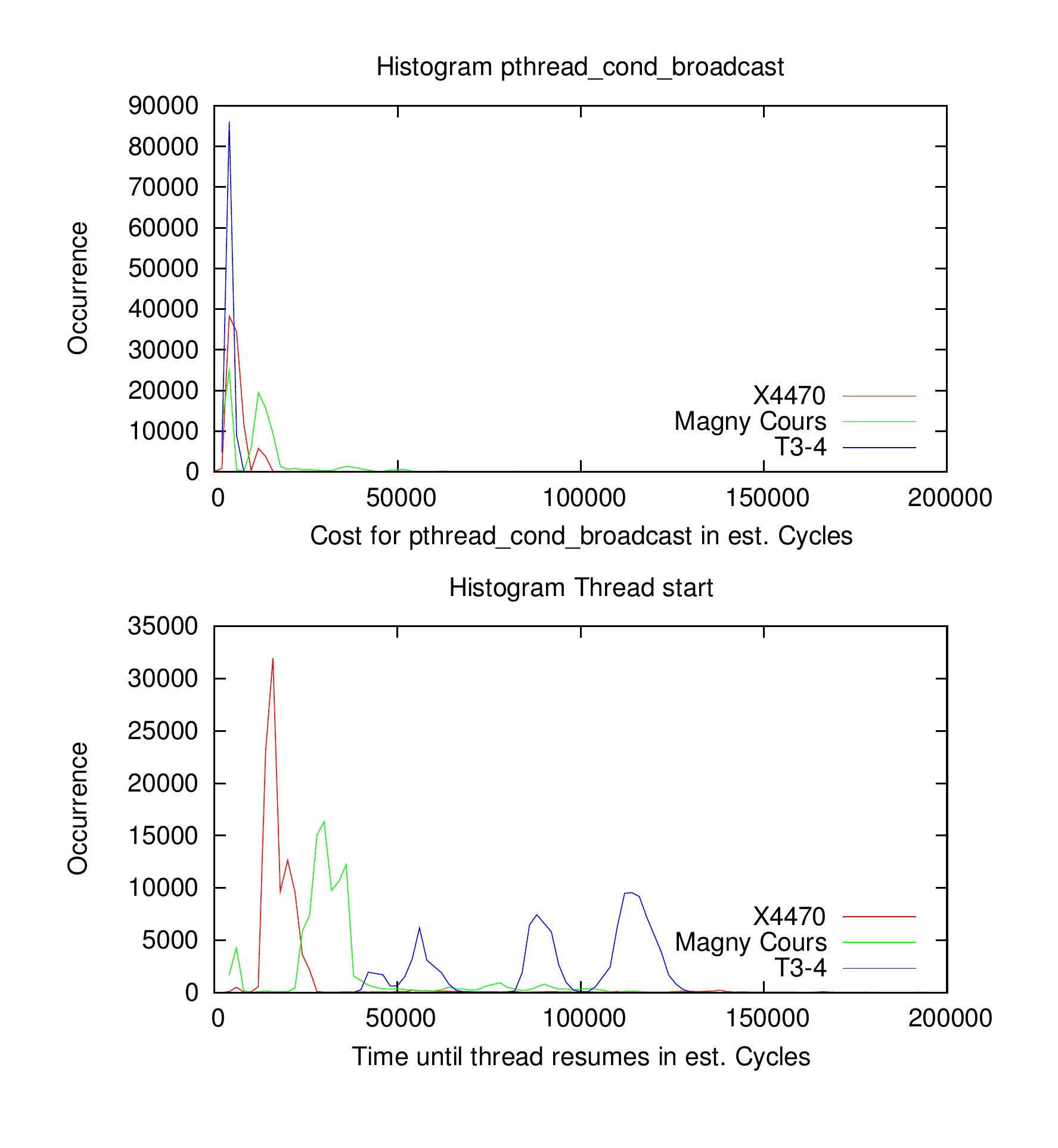}
\par\end{centering}

\begin{centering}
\includegraphics[width=1\textwidth]{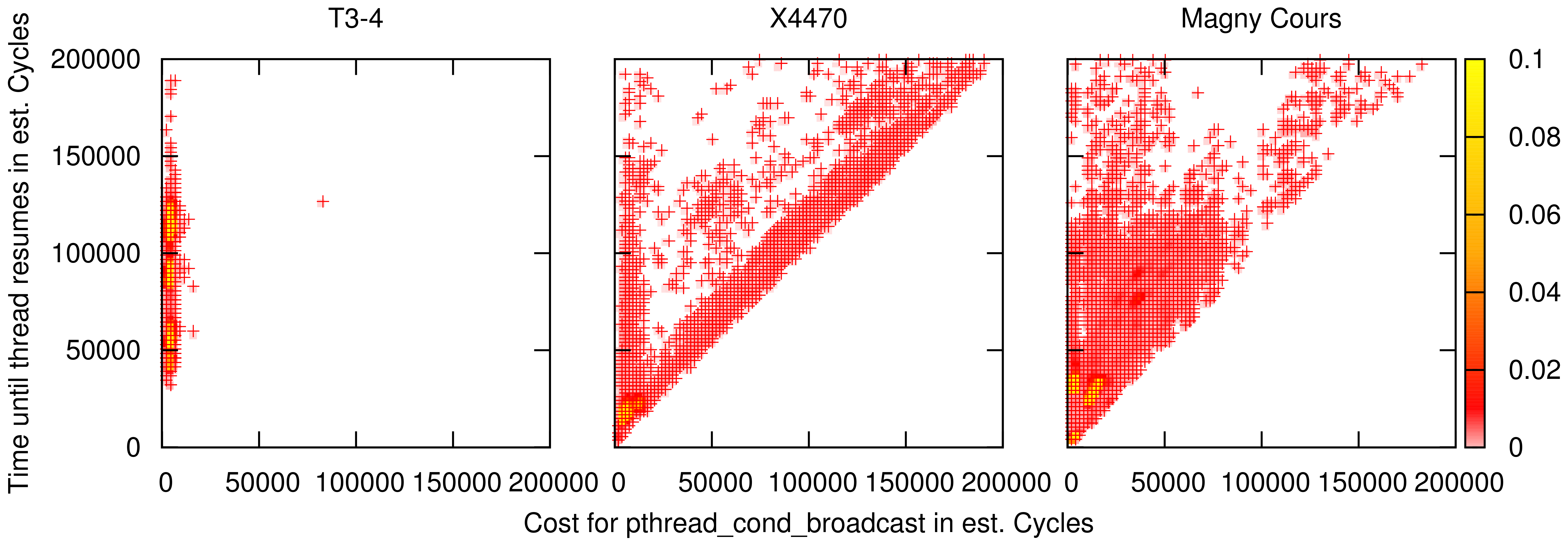}
\par\end{centering}

\caption{\label{fig:cond-bcast-behavior}Measured behavior of \textbf{\emph{pthread\_cond\_broadcast()}}}

\end{figure}
We now look at the results of a similar measurement but using the
\textbf{\emph{pthread\_cond\_broadcast()}} call. Again, the results
for $10^{5}$ measurements are shown in the same way, in \prettyref{fig:cond-bcast-behavior}.
At a first glance, the results are almost the same as those for \textbf{\emph{pthread\_cond\_signal()}},
which is not strange as we only have a single thread waiting on the
conditional. Again, on the \emph{T3-4} 99.9\% of the calls take 2-6
KCycles, however it seems slightly faster on average; 4.7\% at 2 KCycles
versus 0.5\% for the normal signaling, and 9\% versus 15\% at 6 KCycles.
There seems to be no significant difference between the two different
calls on the \emph{X4470}, and there is a slight shift on the \emph{Magny
Cours}; as now 36\% of the calls completed within 2-4 KCycles (18\%
previously), however, less threads started within 4\nobreakdash-8
KCycles, only 5.9\% versus 10.5\%.

\subsection{Conclusions}

Thread creation under Solaris (on SPARC), and therefore also on the
\emph{T3-4}, is still expensive compared to the x86 based Linux systems.
However, mutex synchronization is well optimized and has a very predictable
behavior and outperforms the Linux mutexes roughly in 20\% to 30\%
of the cases. The overhead for signaling or broadcasting a conditional
is also low and predictable on the \emph{T3-4}, outperforming the
\emph{X4470} in 20\% of the cases and the \emph{Magny Cours} in 80\%.
The downside is the longer delay at which the threads are woken up;
this is mainly (99\%) spread between 42 and 124 KCycles on the \emph{T3-4},
which is worse than approximately 93\% of the cases on the \emph{X4470},
and worse than 84\% on the \emph{Magny Cours}.

In order to fully exploit the real hardware thread level parallelism
offered by the \emph{T3-4 }system, it would benefit from more optimized
thread creation and synchronization primitives. When such constructs
cost less overhead, it becomes worthwhile to create more threads of
a smaller granularity. For example, a data-parallel operation could
be split into more smaller threads, able to fill up all the hardware
slots in the \emph{T3-4}. Furthermore, more functional parallelism
could potentially be exploited when the overhead to create it is reduced.
It would also be an advantage to have a scheduling mechanism in user-space
that can schedule work to threads mapped to the cores, for example
like Apple's Grand Central Dispatch \cite{GCD2009}.

\section{SVP-ptl related experiments}\label{sec:SVP-ptl-related-experiments}

After determining the properties of the pthread implementations on
the different systems, we now experiment with the behavior of our
SVP implementation based on pthreads, SVP-ptl \cite{SVP-PTL09}. As
there were some issues with special definitions that the Sun C compiler
did not accept, the code was compiled with GCC \cite{GCC} on all
three machines. On the \emph{T3-4} this was using GCC 4.5.1 with optimization
flags \textbf{\nobreakdash-O2 \nobreakdash-m64 \nobreakdash-mtune=niagara2
\nobreakdash-mcpu=niagara2 \nobreakdash-mvis}, on the \emph{X4470}
and \emph{Magny Cours }we used GCC 4.1.2 with the optimization flags
\textbf{\nobreakdash-O2 \nobreakdash-mfpmath=sse,387 \nobreakdash-mtune=core2
\nobreakdash-mcpu=core2 \nobreakdash-mmmx \nobreakdash-msse \nobreakdash-msse2
\nobreakdash-msse3 \nobreakdash-msse4a} for the \emph{X4470}, and
\textbf{\nobreakdash-O2 \nobreakdash-mfpmath=sse,387 \nobreakdash-mtune=amdfam10
\nobreakdash-march=amdfam10 }for the \emph{Magny Cours} system.

\subsection{FFT}

In this experiment we run an SVP version of the Fast Fourier Transform
algorithm, which was transformed to expose maximum concurrency between
its iterations for execution on the Microgrid, on which it shows a
nice scalability and speedup. The performance figures in this experiment
do not reflect in any way the performance of the FFT algorithm on
the tested systems, but they do show how the SVP-ptl run-time copes
with many small short-lived threads. As creating pthreads is expensive,
as also shown in \prettyref{sec:Pthread-experiments}, the run-time
keeps a pool of pthreads that it assigns work to. When an SVP thread
needs to be created, the runtime first checks the pool for an available
thread, and otherwise starts another pthread. When a pthread finishes
its work, it puts itself in the pool, unless it exceeded the maximum
pool size. We investigate the effect of two different implementations
of this thread pool here. One uses a pthread mutex to guard the pool
so that it can be exclusively updated, and the second is a lockless\emph{
}implementation using an atomic Compare-And-Swap (CAS) instruction
to add/remove entries to the pool.

\begin{figure}
\begin{centering}
\includegraphics[width=0.75\textwidth]{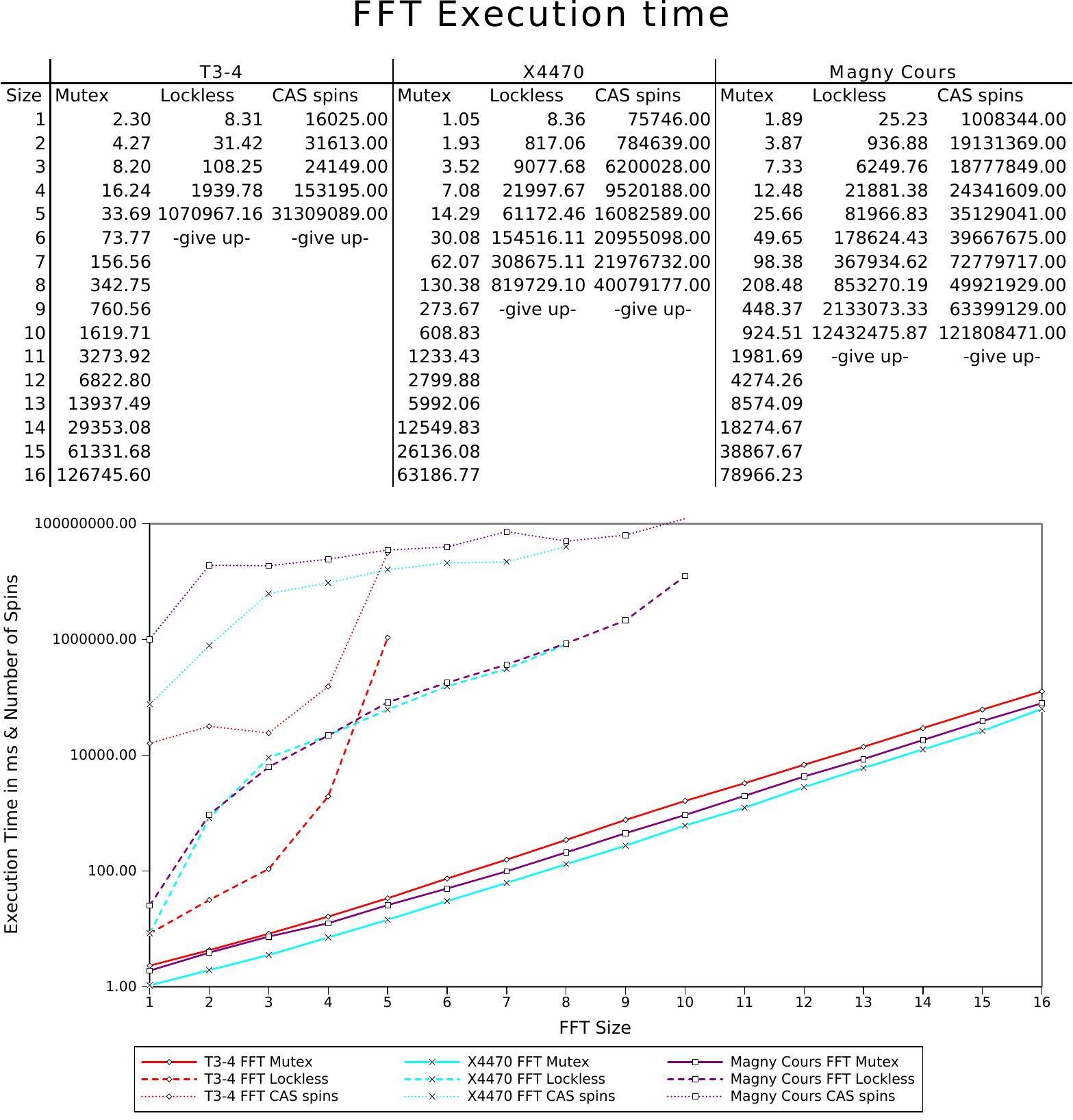}
\par\end{centering}

\caption{\label{fig:FFT-results}Table and plot showing results of FFT experiment}

\end{figure}
For the experiment, we run the FFT implementation with the two different
pool implementations for various FFT sizes varying from 1 to 16, with
the number of created threads growing exponentially for each size.
The results of these experiments are shown in \prettyref{fig:FFT-results},
showing the execution time in ms for both implementations, and the
total number of times the run-time had to spin on a CAS instruction
until it completed successfully. As the execution time (and the amount
of work that needs to be done) grows exponentially with the increase
in size, the y-axis is set to a logarithmic scale.

The first experiment, using the thread pool guarded by a mutex is
shown as {}``FFT Mutex'' in \prettyref{fig:FFT-results}. The \emph{T3-4}
is roughly two times slower than the \emph{X4470}, but this was to
be expected after the results we saw in \prettyref{sec:Pthread-experiments},
showing larger thread creation and synchronization overheads. What
is clear is that all three systems scale up well in this experiment,
roughly doubling their execution time for every increase in FFT size,
as expected.

In the second experiment, we use the thread pool that is updated by
atomic CAS instructions. For the \emph{T3-4} system we use the \textbf{\emph{atomic\_cas\_ptr()}}
function that is defined in \textbf{\emph{atomic.h}}, and on the Linux
systems we use the GCC builtin \textbf{\emph{\_\_sync\_val\_compare\_and\_swap()}}
which is x86 specific. Before we made this experiment we thought it
would perform better, as there would be less contention without a
global lock on the pool. In earlier experiments on single and dual
core systems, this implementation gave us a speedup of around 5\%.
But, as can easily be seen in \prettyref{fig:FFT-results}, this was
not the case at all. The figure shows both the execution time for
different FFT sizes, as well as a count of the total number of times
all threads had to spin on a CAS instruction until they have successfully
made an update on the pool. As this is counted with a non-atomic increment
of a \emph{volatile} global variable, the number is not guaranteed
to be 100\% accurate, but it gives us sufficient insight in what happens.

When we look at the lockless results for the \emph{T3-4} in \prettyref{fig:FFT-results},
we see that the execution time and the number of CAS spins really
explodes. Taking into account that the y-axis in the plot is already
logarithmic, it seems to increase double exponentially. After 5 iterations,
we decided to give up the experiment as the next iteration (if it
would complete at all) would potentially take a whole day or longer,
and the measured data already gave us enough insight in the behavior
on the \emph{T3-4}. We then repeated the experiment on the \emph{X4470}
and the \emph{Magny Cours}, which showed roughly similar behavior,
with the \emph{Magny Cours} requiring more CAS spins. Also both x86
systems did not completely explode in CAS spins as the \emph{T3-4}
did, this is probably because the \emph{T3-4} has many more hardware
threads, and we suspect that at some point up to 512 threads were
spinning on a CAS operation where only one of them would succeed each
time. On the \emph{Magny Cours} this could only be 48 threads, and
on the Hyperthreaded cores of the \emph{X4470} it could be even only
24 threads, explaining the difference between the three systems. Another
reason that makes the situation escalate is that when the run-time
can't get a worker thread from the pool it will start a new pthread,
which after completing it's work will also start to compete with CAS
spins to enter itself into the pool again. This is probably the cause,
together with the massive number of hardware threads, that causes
the observed behavior on the \emph{T3-4}.

Another interesting experience we had during the FFT experiment of
size 5 on the \emph{T3-4} was that the system appeared to completely
hang for a while (around 2 minutes), making us initially think that
we had crashed it. However, it still responded on the network to ping
command's ICMP packets, and after a long delay our remote login shells
started to respond again. Probably the system was completely swamped
with coherency traffic from all the CAS operations, something we did
not experience this heavily on the two other reference systems. Again,
this difference is likely related to the great number of hardware
threads offered on the \emph{T3-4}.

\subsection{Matrix Multiplication}

In the second experiment based on the SVP-ptl implementation, we use
a more suitable granularity of threads for this specific SVP implementation.
We run an SVP implementation of a matrix multiplication using recursive
block decomposition. Both $N\times N$ matrices are divided into four
sub-matrices of $\frac{N}{2}\times\frac{N}{2}$, resulting in eight
concurrent multiplications of these followed by four matrix additions.
Of course the same can be applied recursively to each of the eight
multiplications of sub matrices, yielding $8^{k}$ concurrent activities
where $k$ is the number of recursions. At the inner recursion when
two matrices of $n\times n$ are multiplied, this results in $2\cdot n^{3}$
multiplication and addition operations. As we run the experiments
for $1\ldots5$ recursions and with $N$ ranging from $1024$ to $8192$,
the smallest threads are created in the 5th recursion on a $1024\times1024$
matrix, which still yields $2\cdot(\frac{1024}{2^{5}})^{3}=65536$
floating point operations per thread. The largest threads occur with
a single  recursion on a $8192\times8192$ matrix, which then yields $2\cdot(\frac{8192}{2^{1}})^{3}=1.37\cdot10^{11}$
floating point operations per thread. In this experiment we have the
default mutex based thread pooling enabled, and we leave the mapping
of the threads up to the operating system.

\begin{figure}
\begin{centering}
\includegraphics[width=1\textwidth]{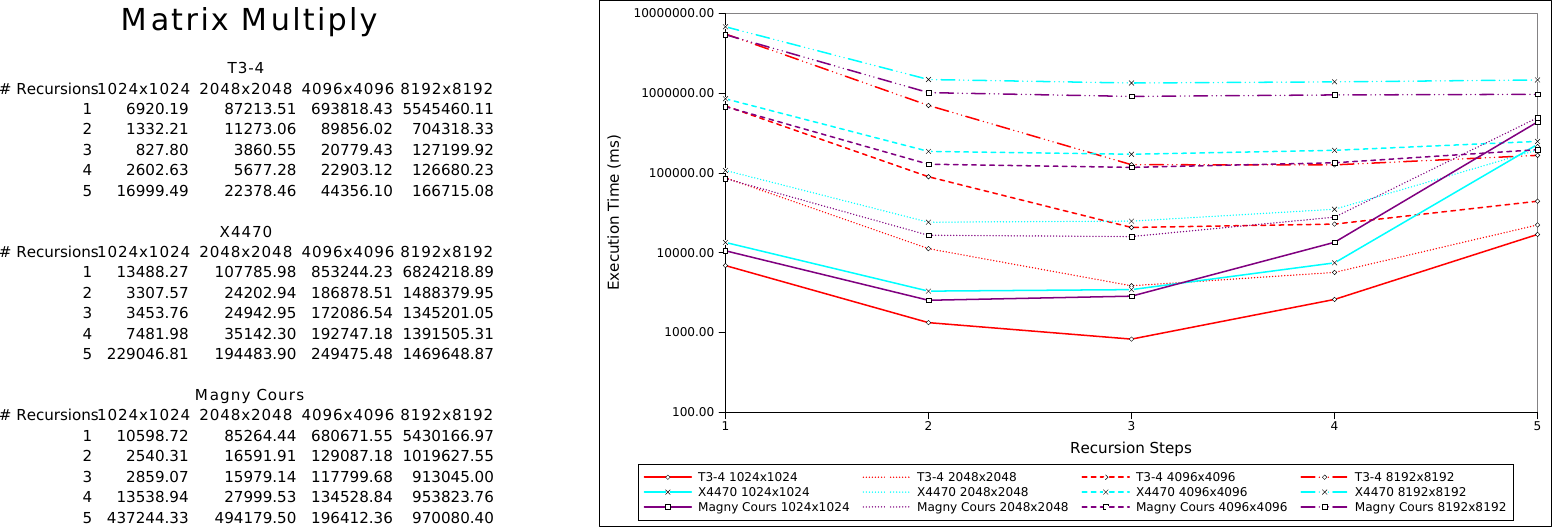}
\par\end{centering}

\caption{\label{fig:Recursive-Matrix-Multiplication}Table and plot showing
results of recursive Matrix Multiplication experiment}

\end{figure}
We ran our matrix multiplication code on all three systems and compare
the execution time for different matrix sizes and number of decomposition
recursions in \prettyref{fig:Recursive-Matrix-Multiplication}. Surprisingly
enough, the \emph{T3-4} outperforms both the \emph{X4470} and \emph{Magny
Cours} for the smallest matrix size, with only one recursion. This
was not what we would expect after our earlier experiments on computational
throughput, where these two systems clearly outperformed the \emph{T3-4}
with 8 computational threads (see \prettyref{fig:Floating-point-Arithmetic-Throughput}).
However, in contrast to that experiment, this has a more regular computation
which can also be expressed as a multiply-accumulate for which the
\emph{T3-4} has a dedicated instruction which is not present on current
x86 architectures. This might explain why the \emph{T3-4} outperforms
the two reference systems for all matrix sizes with a single recursion.

The \emph{T3-4} scales up well to 3 recursions, which is to be expected
as this corresponds with a maximum of $8^{3}=512$ concurrent threads.
The optimum for the two reference systems lies clearly at 2 recursions,
corresponding to a maximum of $8^{2}=64$ threads, which lies the
closest to the number of hardware contexts (48) both systems have.
Interestingly, for large matrix sizes the deeper recursions don't
seem to give much penalty on the reference systems, but for the $1024\times1024$
and $2048\times2048$ matrices this is not the case, and at 5 recursions
(yielding $8^{5}=32768$ threads at a maximum) they suffer a lot from
overhead. The \emph{T3-4} manages this relatively well; 5 recursions
gives a 20-fold slowdown on a $1024\times1024$ matrix compared to
the optimum at 3 recursions, against a factor of 69 on the \emph{X4470},
or 174 on the \emph{Magny Cours}. For the largest matrix multiplication,
$8192\times8192$, we really see the \emph{T3-4} take the advantage
of its larger computational- and memory throughput, which we already
demonstrated in \prettyref{sec:Computing-Throughput-Experiment} and
\prettyref{sec:Memory-Throughput-Experiment}.

\subsection{Conclusions}

The \emph{T3-4} system seems to scale up to a lot of threads very
well, and performs very good in our matrix multiplication experiments
probably due to the hardware multiply-accumulate support. However,
when a lot of threads are using atomic compare-and-swap, it becomes
a nightmare, even worse than the x86 based reference systems. It would
be interesting to investigate if the other atomic operations defined
by \textbf{\emph{atomic.h}} suffer from the same effect, and hence
should be avoided. The advantage of using synchronizations of the
pthread library such as mutexes, is that it will suspend a thread
to the OS scheduler instead of having a thread spin in retries making
the situation worse.

\section{Overall Conclusion}

In this report we have tried to explore the \emph{T3-4 }system from
our own research's perspective, and we compared it to two other x86-based
multi-core systems that we had available. We have investigated the
scalability limits of both raw computing and memory throughput, where
the \emph{T3-4} showed very good scalable performance which amazed
us. It really takes 512 threads to saturate the integer or floating
point arithmetic units, and up to 256 threads to saturate the memory
read bandwidth which outperforms the reference systems by a factor
two. Especially that last fact helps the architecture to live up to
the promises about CMT in the T3 Whitepaper \cite{T3whitepaper2010};
as plenty of memory bandwidth is required to feed the many threads
of execution. The experiments confirmed these nice dataflow like scheduling
properties between threads in one core, with very fine-grained hardware
multithreading and latency hiding.

Our experiments with mapping threads to specific cores showed that
Solaris is well capable of distributing the work across all cores
in a nearly optimal way by itself. We also revealed that creation
of threads and synchronizations are still expensive in Solaris, and
are outperformed by the x86-based reference systems that run Linux.
However, the \emph{T3-4} seemed to deal much better with extreme amounts
of threads than the other two reference systems. Finally, we discovered
that atomic operations can be very tricky on such a system with that
many threads actually executing in hardware in parallel, possibly
ending up in nightmare scenarios as we had the \emph{T3-4} system
hanging for 2 minutes. The x86 based systems showed similar problems,
but due to the much lower number of hardware contexts, (more than
a factor of 10), the problem was not as prevalent.

\bibliographystyle{ieeetr}
\addcontentsline{toc}{section}{\refname}\bibliography{references}

\section*{Acknowledgement}\addcontentsline{toc}{section}{Acknowledgement}
The author would like to thank Sun/Oracle for early access to a SPARC T3-4 system through the CMT beta programme which made these experiments possible, as well as the feedback given on the work and this report.

\section*{About the Author}\addcontentsline{toc}{section}{About the Author}

drs. Michiel W. van Tol is currently a PhD student in his final year at the Computer Systems Architecture group at the University of Amsterdam, where he
also received his MSc degree in 2006, specializing in Computer Architecture.
He has a broad background and interest in computer systems, including
nearly ten years of experience as a professional Linux System Administrator.
His research interests include Many-core Architectures, Operating
Systems, Parallel run-times and Multi-threading, Distributed Systems
and Resource Management. Currently he is writing his PhD thesis on
designing Operating Systems for SVP based Many-core architectures.

\end{document}